\documentclass[useAMS,usenatbib,structabstract]{mn2e}

\usepackage{graphicx,latexsym}
\usepackage{color}
\usepackage{natbib}
\usepackage{rotating,ams}
\usepackage[ps2pdf,linktocpage]{hyperref}
\hypersetup{%
  colorlinks=false,
  bookmarksopen=true,
  bookmarksnumbered=true,
  pdfstartpage={1},  
  pdftitle = {How well do {\sc starlab} and {\sc nbody4} compare? I: Simple models},
  pdfauthor = {P. Anders, H. Baumgardt, N. Bissantz, S. Portegies Zwart} ,
  pdfkeywords = {N-body simulations},
  pdfproducer = {LaTeX with hyperref}
}

\def\spose#1{\hbox to 0pt{#1\hss}}
\def\lta{\mathrel{\spose{\lower 3pt\hbox{$\mathchar"218$}}
     \raise 2.0pt\hbox{$\mathchar"13C$}}}
\def\gta{\mathrel{\spose{\lower 3pt\hbox{$\mathchar"218$}}
     \raise 2.0pt\hbox{$\mathchar"13E$}}}

\vspace{-1.5cm}
\title[Comparison NBODY4 vs STARLAB I.]{How well do {\sc
starlab} and {\sc nbody4} compare? I: Simple models}

\author[P. Anders et al.]{P. Anders $^{1}$ \thanks{E-mail: P.Anders@uu.nl}, 
H. Baumgardt $^{2}$, N. Bissantz $^{3}$, S. Portegies Zwart $^{4,5}$\\
$^{1}$ Sterrenkundig Instituut, Universiteit Utrecht, P.O. Box 80000, 
NL-3508 TA Utrecht, The Netherlands\\
$^{2}$ Argelander Institut f\"ur Astronomie, Universit\"at Bonn, Auf dem
H\"ugel 71, 53121 Bonn, Germany\\
$^{3}$ Fakult\"at f\"ur Mathematik, Ruhr-University of Bochum, Mathematik
III, NA 3/70, Universit\"atsstra{\ss}e 150, 44780 Bochum, Germany\\
$^{4}$ Astronomical Institute `Anton Pannekoek' University of Amsterdam,
The Netherlands\\
$^{5}$ Section Computational Science, University of Amsterdam, The
Netherlands
}
\begin{document}

\date{Accepted ---. Received ---; in original form ---.}
\pubyear{2008}
\maketitle

\begin{abstract}
N-body simulations are widely used to simulate the dynamical evolution
of a variety of systems, among them star clusters. Much of our
understanding of their evolution rests on the results of such direct
N-body simulations. They provide insight  in the structural evolution of
star clusters, as well as into the  occurrence of stellar exotica.
Although the major pure N-body codes STARLAB/KIRA and NBODY4 are widely
used for a range of applications, there is no thorough comparison study
yet.

Here we thoroughly compare basic quantities as derived from simulations
performed either with STARLAB/KIRA or NBODY4.

We construct a large number of star cluster models for various stellar
mass function settings (but without stellar/binary evolution, primordial
binaries, external tidal fields etc), evolve them in parallel with
STARLAB/KIRA and NBODY4, analyse them in a consistent way and  compare
the averaged results quantitatively. For this quantitative comparison we
develop a bootstrap algorithm for functional dependencies.

We find an overall excellent agreement between the codes, both for the
clusters' structural and energy parameters as well as for the properties
of the dynamically created binaries.  However, we identify small
differences, like in the energy conservation before core collapse and
the energies of escaping stars, which deserve further studies.

Our results reassure the comparability and the possibility to combine
results from these two major N-body codes, at least for the purely
dynamical models (i.e. without stellar/binary evolution) we performed.
Further detailed comparison studies for more complex systems, e.g.
including stellar/binary evolution, are required.

\end{abstract}

\begin{keywords} 
Methods: N-body simulations, Methods: statistical, open clusters and
associations: general
\end{keywords} 

\section{Introduction}

In recent years, stellar dynamics has led to an advance in a  variety of
fields, such as studies of individual star clusters
(\citealt{2005MNRAS.363..293H}),  star cluster systems
(\citealt{2003ApJ...593..760V}), populations of ``exotic'' objects in 
star clusters (e.g. runaway merger products with masses of up to few
thousands solar masses, \citealt{2004Natur.428..724P}; blue  stragglers,
\citealt{2005MNRAS.363..293H,2007MNRAS.374...95P}), the formation and
evolution of  higher-order hierarchical systems (triples, quadruples and
higher, e.g. \citealt{2007MNRAS.379..111V}), the Galactic centre  and
runaway stars
(\citealt{2005MNRAS.363..223G,2006MNRAS.372..174B,2008MNRAS.384..323L}),
etc. 

The major codes used in this field are the family of {\sc nbodyx} codes
(\citealt{1999PASP..111.1333A}; the most widely used versions are {\sc
nbody4, nbody6, nbody6++}, the most recent version being {\sc nbody7})
and the {\sc starlab} environment with its N-body integrator {\sc kira}
(\citealt{2001MNRAS.321..199P}). Despite these codes being widely used,
there is no thorough comparison study yet. First attempts have been
initiated by Douglas Heggie and others at the IAU General Assembly 1997
in Kyoto (therefore, the ``Kyoto experiment''), however until today, the
number of results and their analysis is small (see
\citealt{2001astro.ph.10021H} and
http://www.maths.ed.ac.uk/$\sim$heggie/kyotoII/kyotoII.html for
descriptions of this collaborative experiment and some of its results).

Although the fundamental integration scheme (4$^{th}$ order,
block-timestep ``Hermite'' predictor-corrector scheme, see
\citealt{1992PASJ...44..141M}, the next section in this paper, or
\citealt{2003gnbs.book.....A} for a variety of technical details) is the
same for both codes, severe differences in the treatment of binaries,
stellar and binary evolution are present, plus naturally different
implementations of otherwise comparable components.

With this paper we start a series of publications to study the impact
of  differences in the input physics onto the results from both codes.
We  start with the most simple models, not including stellar evolution,
external tidal fields or primordial binaries. More complex models,
without the  aforementioned restrictions, will be studied in upcoming
papers of this series.

In these studies we concentrate on the {\sl statistical} treatment of a
large number of runs, represented by its median values and uncertainty
ranges, as the results of single runs will naturally diverge due to the
amplification of numerical errors (\citealt{1993ApJ...415..715G}). This
holds for two models with slightly different initial configurations and
evolved with the same code, as well as the same initial model evolved
with two different codes. We will not compare wall-clock times for the
different runs (as this is dependent on a multitude of parameters,
software and hardware settings), or parameters usually only relevant to
code developers. Instead we want to provide the interested scientific
user a guideline to the comparability of the codes studied and point at
differences concerning parameters likely relevant for the user.

The paper is organised as follows: Sect. \ref{sec:physics} gives an
overview of similarities and differences of the {\sc starlab} and {\sc
nbody4} input physics. Sect. \ref{sec:setup} describes the general model
setup and the data analysis pipeline. In Sect. \ref{sec:bootstrap} we
introduce a bootstrap approach to quantify differences between
functions. In Sect. \ref{sec:results} we present our results for a range
of mass function settings. Energy conservation is studied in Sect.
\ref{sec:resEC}, core collapse in Sect. \ref{sec:cc}, and the properties
of stars becoming unbound during cluster evolution in Sect.
\ref{sec:escapers}. We finish this paper with our conclusions in Sect.
\ref{sec:conclusions}.

\section{Overview}
\label{sec:physics}

\subsection{Similarities in input physics: N-body integrator scheme}

Almost all recent direct N-body integrators (including {\sc nbody4} and
{\sc kira/starlab}) are based on the 4$^{th}$ order, block-timestep
``Hermite'' predictor-corrector scheme (but see e.g.
\citealt{2008NewA...13..498N} for higher-order N-body integrators).

{\sl ``Hermite'' predictor-corrector scheme}: This integration scheme
was first described by \citet{1991ApJ...369..200M}. It is based on
individual timesteps for every star, (approximate) prediction of all
stars' positions and derivatives, and (accurate) calculation and
correction (using Hermite interpolation) for a subset of stars at any
given timestep.

More specifically, the scheme comprises of the following steps to evolve
one star $i$ from the present time $t_0$ to the time $t_1 = t_0 + \Delta
t$. The positions, velocities, acceleration and the first time
derivative of the acceleration at time $t_0$ are assumed to be known for
all stars. This description follows \citet{1992PASJ...44..141M}. 

\begin{enumerate}

\item Predict/extrapolate the positions, velocities, accelerations and
the first time derivatives of the acceleration at time $t_1$, using
Taylor expansion (up to 4$^{th}$ order for the positions) with the
quantities at time $t_0$, for all stars.

\item Calculate for star $i$ the acceleration and the first time
derivative of the acceleration at time $t_1$, based on the predicted
positions and velocities of all stars.

\item Calculate for star $i$ the second and third time derivatives of
the acceleration at time $t_0$, using the acceleration and the first
time derivative of the acceleration at times $t_0$ and $t_1$.

\item Calculate for star $i$ the correction to the predicted position
and velocity, based on the second and third time derivatives of
the acceleration at time $t_0$.

\end{enumerate}

{\sl Block timesteps}: In principle, the optimum timestep can be estimated for
each star individually, based on this star's acceleration and its time
derivatives. In reality, it is computationally favourable to group stars
with approximately the same timestep together, and evolve whole
``blocks'' of stars at once. This treatment reduces the overheads
otherwise needed to calculate the predicted positions and velocities of
all stars. Conventionally, block timesteps of power-of-2 ($\Delta t_n
\propto 1/2^n$) are used.

\subsection{Differences in input physics: Treatment of binaries}

The treatment of binaries is one of the challenges in N-body
simulations. While especially close binaries are dynamically important
(e.g. star-binary and binary-binary interactions can eject stars from
the cluster core, resulting in the halting of core collapse and leading
to core re-expansion), their relevant timescale is the orbital period
(of the order of days), while the relevant timescale for the cluster as
a whole is the crossing timescale (of the order of Myrs). A ``brute
force'' approach would need to set the timestep to a fraction of the
orbital period to evolve the binary accurately, which would immediately
stall the calculation (and corrupt energy conservation due to
exponential growth of numerical inaccuracies).

However, especially close/hard binaries are hardly perturbed by external
effects, as their binding energy is high compared to the energy injected
by external perturbations. Such binaries evolve essentially as in
isolation.

{\sc nbody4} uses the KS (\citealt{KS}) and CHAIN
(\citealt{1993CeMDA..57..439M}) regularisation techniques to follow
close encounters between stars. The basic idea of these regularisation
methods is to switch to special coordinate systems together with
appropriate time transformations which significantly improve the overall
energy conservation during close encounters. 

%

{\sc starlab} separates between ``unperturbed/hard binaries'' and
``perturbed binaries'', where the distinction is made where the
dimensionless perturbation (i.e. the ratio of the external perturbation
to the internal binary binding energy) reaches a critical value
(typically 10$^{-6}$). Unperturbed binaries are evolved solving
analytically the Kepler equations, and their components are treated as
point masses, for the purpose of influencing other stars.

``Slightly perturbed'' bound pairs with a dimensionless perturbation
between 10$^{-6}$ and 10$^{-5}$ are treated with a ``slowdown''
algorithm similar to the one described in \cite{1996CeMDA..64..197M}.

More strongly perturbed pairs and multiples are treated as resolved into
their components for the purpose of determining their influence of
surrounding stars. Their motion is calculated directly.

Perturbations are followed efficiently by keeping a perturber-list for
each binary.

More detailed information are given in \cite{2001MNRAS.321..199P}

\section{Model setup and data analysis}
\label{sec:setup}

In order to simplify future comparisons with other N-body codes, we
provide the input snapshots (both in {\sc starlab/kira} as well as in
{\sc nbody} format),the time evolutions of important cluster parameters,
{\bf and for each code an example run parameter file} on our webpage
(http://www.phys.uu.nl/$\sim$anders/data/ NBODY\_STARLAB\_Comparison/
and http://members.galev.org/nbody/ NBODY\_STARLAB\_Comparison/).

We created 50 input models per setting, using the appropriate {\sc
starlab} tasks, to improve the statistics. For {\sc nbody4}, these
models were converted into the appropriate input format, to have 
maximum comparability.

{\bf All input models were evolved for 1000 N-body time units, well
beyond core collapse, using {\sc nbody4} (the May 2008 version from
Aarseth's
webpage\footnote{http://www.ast.cam.ac.uk/~sverre/web/pages/nbody.htm})
respectively {\sc starlab/kira} (throughout the paper we use {\sc
starlab} version 4.4.2). All simulations were performed on PCs hosting
GRAPE special-purpose hardware (see e.g. \citealt{2003PASJ...55.1163M};
the {\sc nbody4} runs were performed on a machine hosting a GRAPE6A
board, the {\sc starlab} on a machine hosting a GRAPE6BLX board). As
{\sc nbody6} does not support the usage of GRAPE hardware, we limit our
analysis to {\sc nbody4} in the course of this paper. For the impact of
the hardware (and the associated internal calculation accuracy) on the
results of our N-body simulations, see \citet{anders08}. Preliminary
results indicate little impact.}

Crucial for our comparison is a self-consistent analysis of the  {\sc
nbody4} and {\sc starlab/kira} output. In order to achieve this goal we
convert the {\sc starlab} output into the {\sc nbody4} output format
(with the same number of significant digits = 15), removing all
information not available for the {\sc nbody4} output, like local
densities, binary parameters, energies, cluster centres etc. {\sl At
this stage the binary tree structure is not yet established and the
binary/multiple components are treated as single stars.}

The results, snapshot by snapshot, were then fed into {\sc
starlab/kira}  and evolved for a short time (a 1/32$^{th}$ of an N-body
time unit), {\sl in order to reconstruct the binary populations}. From
the resulting snapshots we calculate a large number of
diagnostics\footnote{We used the analysis task {\sc hsys\_stats} in {\sc
starlab}, and recalculate energies, core radii and other quantities
requiring O(N$^2$) operations where necessary.}: structural parameters
(cluster centres, mass profile, Lagrange radii, King parameter, core
density etc), energies (potential energy, kinetic energy, energy error
etc), and parameters of dynamically created binaries (eccentricities,
binding energies, positions of the binaries inside the cluster etc).

For the majority of this paper we will concentrate on the results from
snapshots with fully reconstructed binary tree structure. The impact of
the binary tree reconstruction on the data will be discussed in Sect.
\ref{sec:reswobin}.

\subsection{Nomenclature}
\label{sec:nom}

We will use the following definitions and abbreviations throughout the
paper.

\begin{itemize}

\item {\sl standard runs = std}: simulations made with the standard 
settings described in Sect. \ref{sec:stddef}

\item {\sl MF10 runs}: simulations made with the standard settings
described below, except that a \cite{1955ApJ...121..161S} mass function is used,
with the upper mass limit being 10x larger than the lower mass limit

\item {\sl MF100 runs}: simulations made with the standard settings
described below, except that a \cite{1955ApJ...121..161S} mass function is used,
with the upper mass limit being 100x larger than the lower mass limit

\item {\sl unperturbed binaries}: relative external perturbation is
smaller than 10$^{-6}$

\item {\sl perturbed binaries}: relative external perturbation is larger
than 10$^{-6}$, but binding energy $\arrowvert {\rm E}_{\rm bind}
\arrowvert \ge 0.5$ kT.

\item {\sl multiples}: second strongest bound orbit in a multiple system
(primary mass = total mass of inner binary with strongest bound orbit)

\item {\sl significance level} of statistical test results:
\begin{itemize}
\item {\sl highly significant}: p-value $<$ 1\%
\item {\sl significant}: p-value $<$ 5\%
\item {\sl weakly significant}: p-value $<$ 10\%
\end{itemize}

\end{itemize}

In the remainder of this work, for studying structural parameters and
energies we will use only the median cluster parameters calculated from
the individual runs. The associated uncertainty ranges are the 16\%/84\%
quantiles (similar to the 1$\sigma$ ranges for Gaussian-distributed
quantities around their mean value), divided by the square-root of the
number of runs contributing, to estimate the uncertainty in the position
of the median value. For studying the properties of dynamically created
binaries, we add up all binaries from the individual runs which are
present at a given time.

This procedure reduces the noise from the individual runs. In addition,
as runs inevitably diverge (either two start models with slightly
different initial conditions evolved with the same code, or one start
model evolved with two different codes), a direct comparison of
individual runs is not expected to give meaningful results.

\subsection{Benchmark tests: The ``standard runs''}
\label{sec:stddef}

The benchmark test settings (further on referred to as
``standard runs'') we propose are the following:
\begin{enumerate}
\item 1024 (=1k) particles
\item equal mass system
\item no primordial binaries
\item no stellar/binary star evolution
\item \citet{1911MNRAS..71..460P} sphere density profile
\item no external tidal field
\end{enumerate}

These settings constitute the most simple configuration, which is
likely available  for testing in every future N-body code.

In further sections/papers, several of the restrictions imposed to
establish the  benchmark test settings are going to be dropped, in order
to get more realistic cluster models.

\section{Difference of functions \& Bootstrap test}
\label{sec:bootstrap}

In our study we will obtain time evolutions of various parameters, as
computed for a variety of settings. We want to compare these time
evolutions quantitatively and determine the statistical significance of
differences. For the former one we define a measure how different two
functional relations are, for the latter one we introduce a version of
bootstrapping, adapted to such functional dependencies.

\subsection{Difference of functions: The distance measure}

Assume we have two functional dependencies of one parameter from the
independent variable $x$: $y_1(x)$ and $y_2(x)$. For each $x$ these
dependencies have uncertainties $\sigma_1(x)$ and $\sigma_2(x)$. For
example, from our studies, this relates to $y_1(x) =
r_{core}^{NBODY}(time)$ and $y_2(x) = r_{core}^{STARLAB}(time)$ (i.e.
the core radius at a given time, as determined from {\sc nbody}
respectively {\sc starlab} simulations) and the related uncertainties.

If the data have asymmetric error bars $\sigma_1^+(x)$ and
$\sigma_1^-(x)$, e.g. originating from the use of quantiles, we suggest
to use an average $\sigma_1(x) = 0.5 \cdot (\sigma_1^+(x) +
\sigma_1^-(x))$. However, other measures are possible and, as long as
consistency is ensured, should give similar results.

The relative difference between the functional dependencies at a given
$x$ is then:

\begin{equation}
\delta_{12}(x) = \frac{y_1(x) -
y_2(x)}{\sqrt{\sigma_1(x)^2+\sigma_2(x)^2}}
\label{eq:single}
\end{equation}

We then define the ``difference between functions 1 and 2'' as 

\begin{equation}
\Delta_{12} = \frac{1}{N} \cdot \left | \sum_x \delta_{12}(x) \right |
\label{eq:sum1}
\end{equation}

where $N$ is the number of datapoints used for the statistic. We
consider only the absolute value, as we want to have a measure of the
{\sl size} of the difference, but not necessarily its {\sl direction}.
In addition, this ensures $\Delta_{12} \equiv \Delta_{21}$.

Equivalently we define the ``{\sl absolute} difference between
functions 1 and 2'' as 

\begin{equation}
\Gamma_{12} = \frac{1}{N} \cdot  \sum_x \left | \delta_{12}(x) \right |
\label{eq:sum2}
\end{equation}

While $\Delta_{12}$ is more sensitive to systematic offsets,
$\Gamma_{12}$ traces also statistical fluctuations.

\subsection{Bootstrap test for comparing functions}

We calculate 3 $\times$ 300 test clusters with {\sc starlab} using the
same analysis routines as for the other clusters. These clusters follow
the same settings as the main simulations (i.e. 300 clusters for each
respective mass function).

From these test clusters we randomly select sets of 50 clusters each
(i.e. the number of clusters in the main simulations) {\bf with
replacement}, and calculate for each parameter the median $y^T(x)$ and
quantiles $\sigma^T(x)$. 

We build 2000 such sets. Out of those we randomly select two sets (again
with replacement) and derive the individual values of $\Delta_{12}^T$
and $\Gamma_{12}^T$. We repeat this procedure 10000 times to estimate
the $\Delta_{12}^T$ and $\Gamma_{12}^T$ test distributions for each
parameter. As all test clusters are calculated with the same settings,
the $\Delta_{12}^T$ and $\Gamma_{12}^T$ test distributions represent the
null hypothesis ``functions 1 and 2 are drawn from the same parent
distribution''. By comparing these test distributions with the values
derived from the main simulations $\Delta_{12}^S$ and $\Gamma_{12}^S$ we
can quantify the fraction of data in the test distribution with
$\Delta_{12}^T$ or $\Gamma_{12}^T$ more deviating than the values
derived from the main simulations $\Delta_{12}^S$ or $\Gamma_{12}^S$.
This value serves as measure of how similar the two main simulations
are.

In order to evaluate if the 300 test clusters were sufficient, we
performed the same analysis with a subset of 250 test clusters for the
``MF10'' setting. Depending on the parameter studied, the resulting
comparability p-values differ on average by $\pm$1\% up to maximum
deviations of $\pm$3\%. None of these differences changed the
significance level of any of the results, though.

In order to avoid applying the test statistic to highly correlated data,
which appears for the earliest timesteps (as the {\sc nbody4} and {\sc
starlab/kira} runs share the same start models) and which is beyond the
area of application of the test statistic, we start the summation in Eq.
\ref{eq:sum1} and \ref{eq:sum2} at 10 N-body time units. This value is a
compromise between avoiding early correlated data and containing the
core collapse phase for all models. We tested our method with a range of
starting times and find in general very good agreement. On average
differences are $\sim$3\%, for few extreme cases up to $\sim$15\%, with
a trend of increasing offsets with increasing starting times. This
changes only occasionally the classification of the test result into the
significancy level categories defined in Sect. \ref{sec:nom}, mainly in
cases where the p-value already is close to a boundary between such
categories.

\section{Results}
\label{sec:results}
\subsection{Results for the ``standard runs''}
\label{sec:std}

\begin{figure*}
\begin{center}
  \vspace{-0.5cm}
  \hspace{1.2cm}
  \begin{tabular}{cc}	
	\includegraphics[angle=270,width=0.4\linewidth]{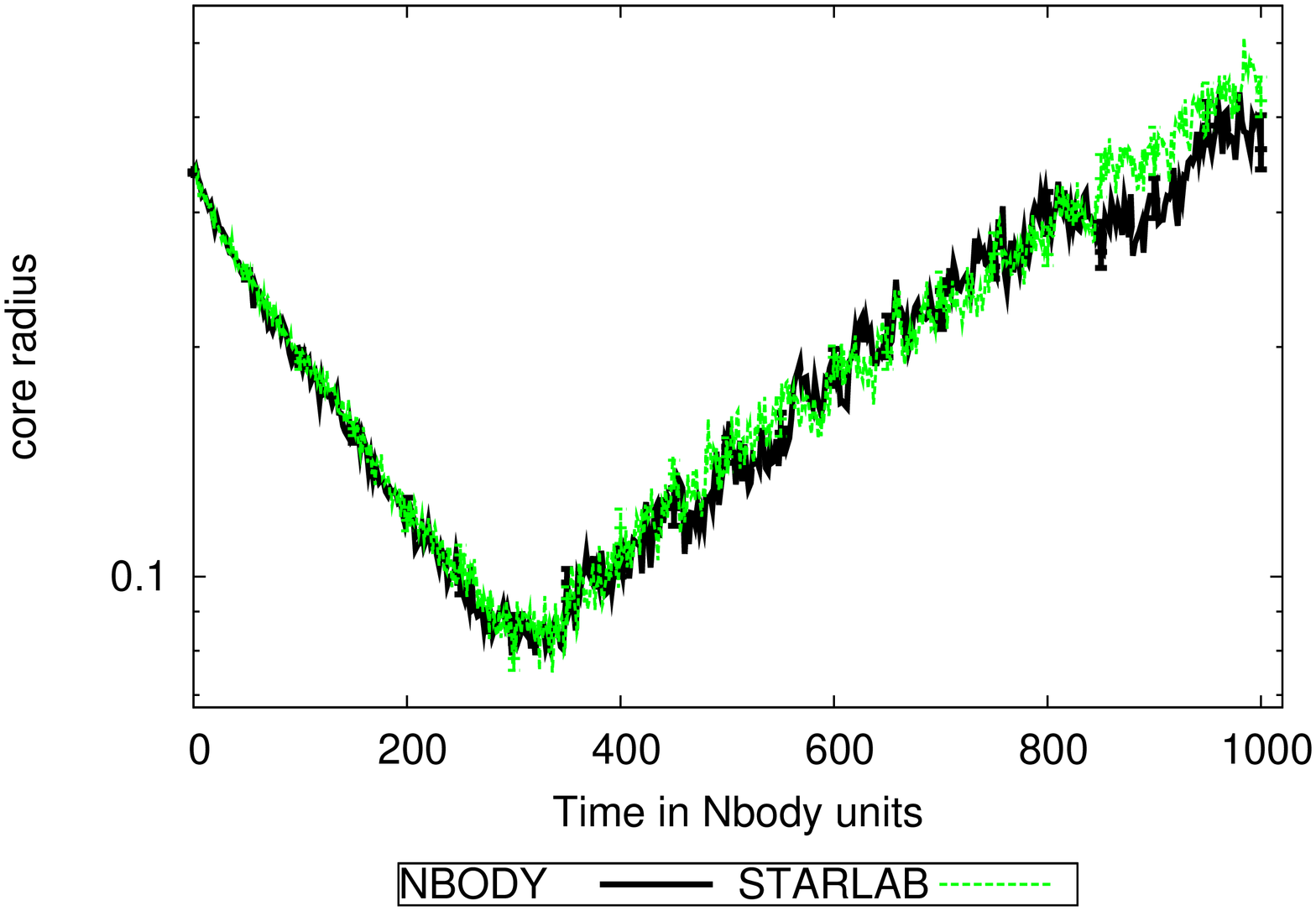} & 
	\includegraphics[angle=270,width=0.4\linewidth]{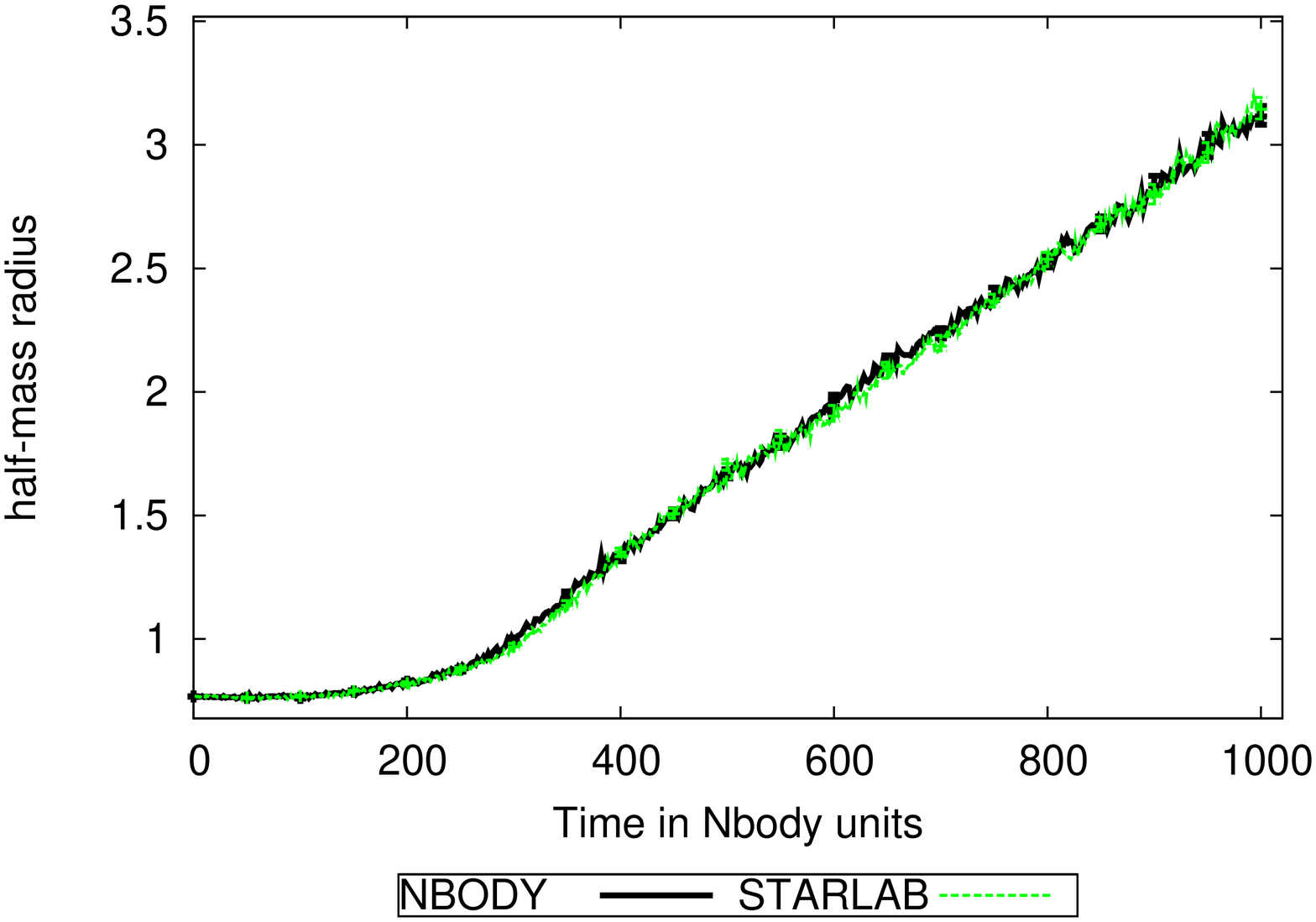} \\
	 \includegraphics[angle=270,width=0.4\linewidth]{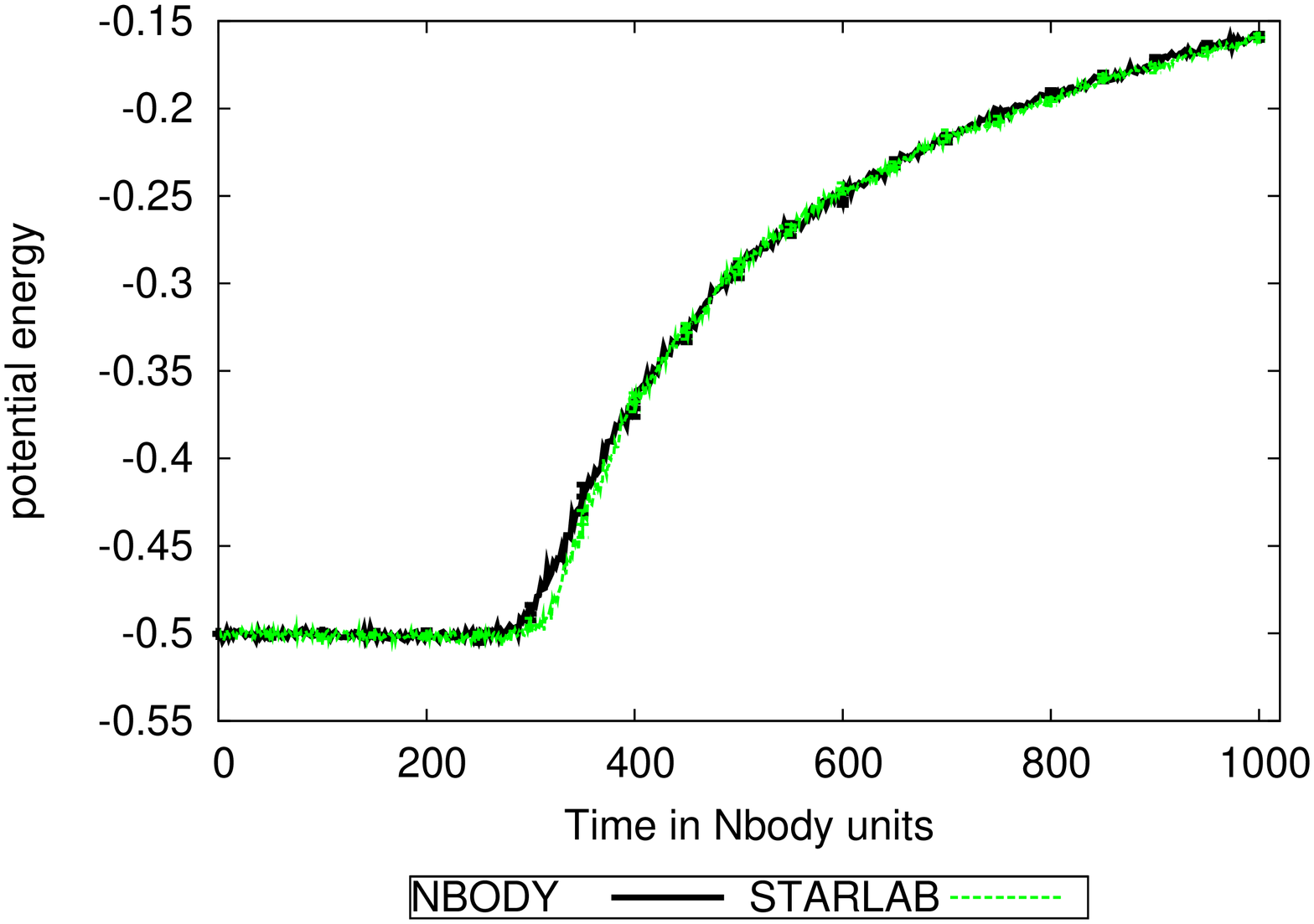} &
	 \includegraphics[angle=270,width=0.4\linewidth]{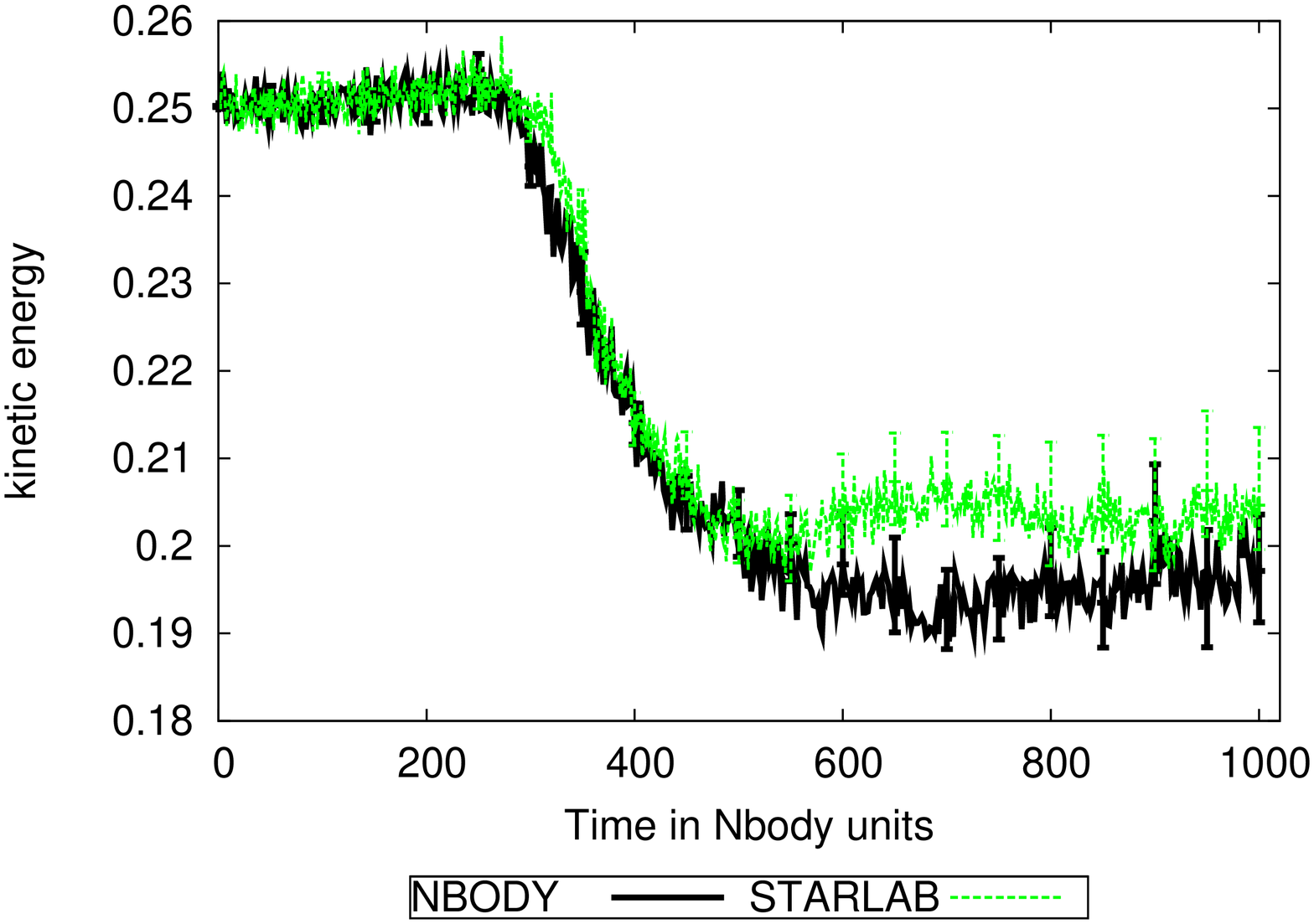} \\  
  \end{tabular}
\end{center}

\caption{Comparison of ``standard runs'' simulations using {\sc starlab}
(green/grey) vs {\sc nbody4} (black).  The lines show the median values,
the error bars give the uncertainty ranges from the 50 individual runs.
Shown are the time evolutions of the core radius (top left),
half-mass radius (top right), potential
energy (bottom left) and kinetic energy (bottom right).}

\label{fig:std_struct_par}

\begin{center}
	\begin{tabular}{cc}
		\includegraphics[angle=270,width=0.4\linewidth]{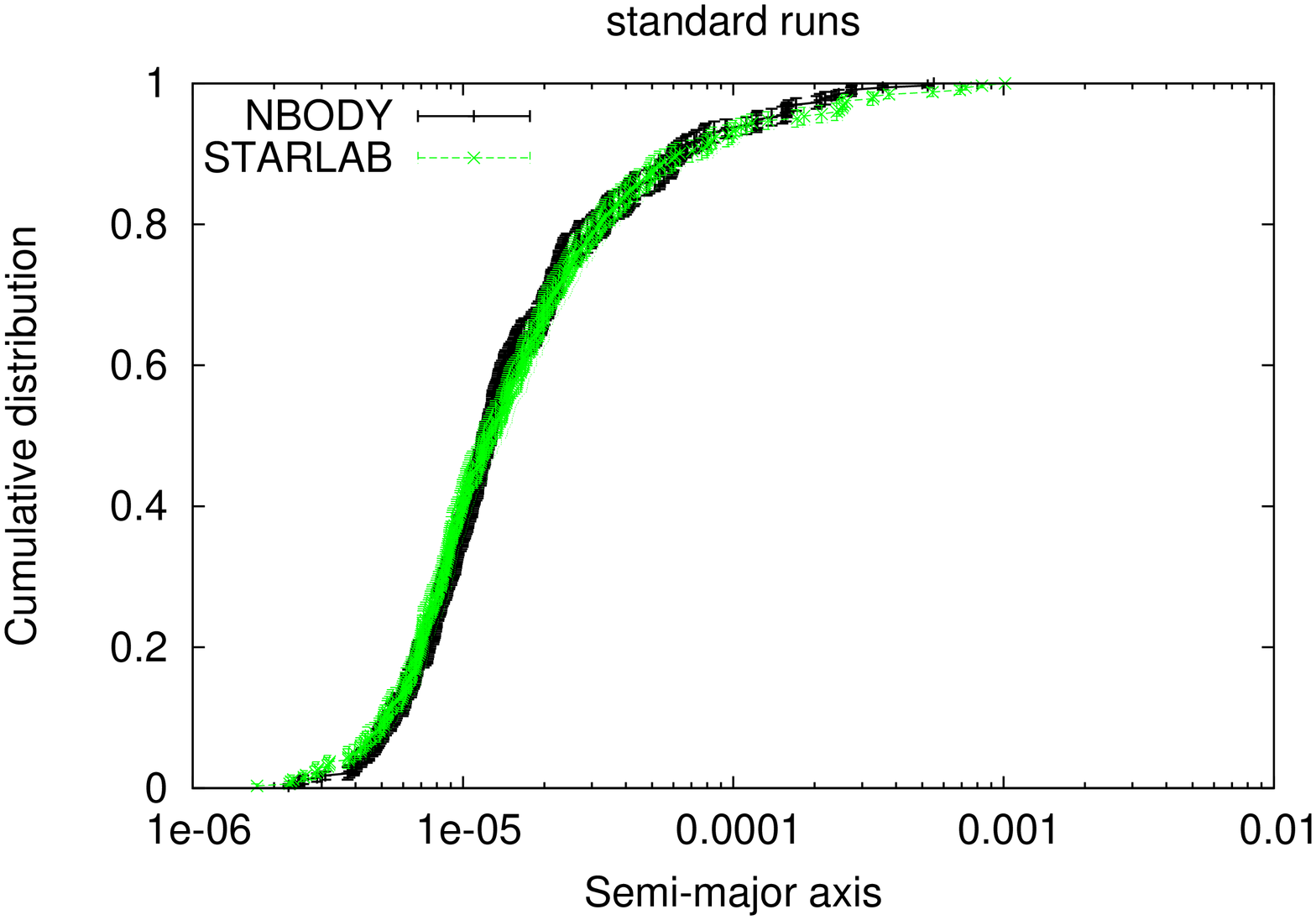} &
 		\includegraphics[angle=270,width=0.4\linewidth]{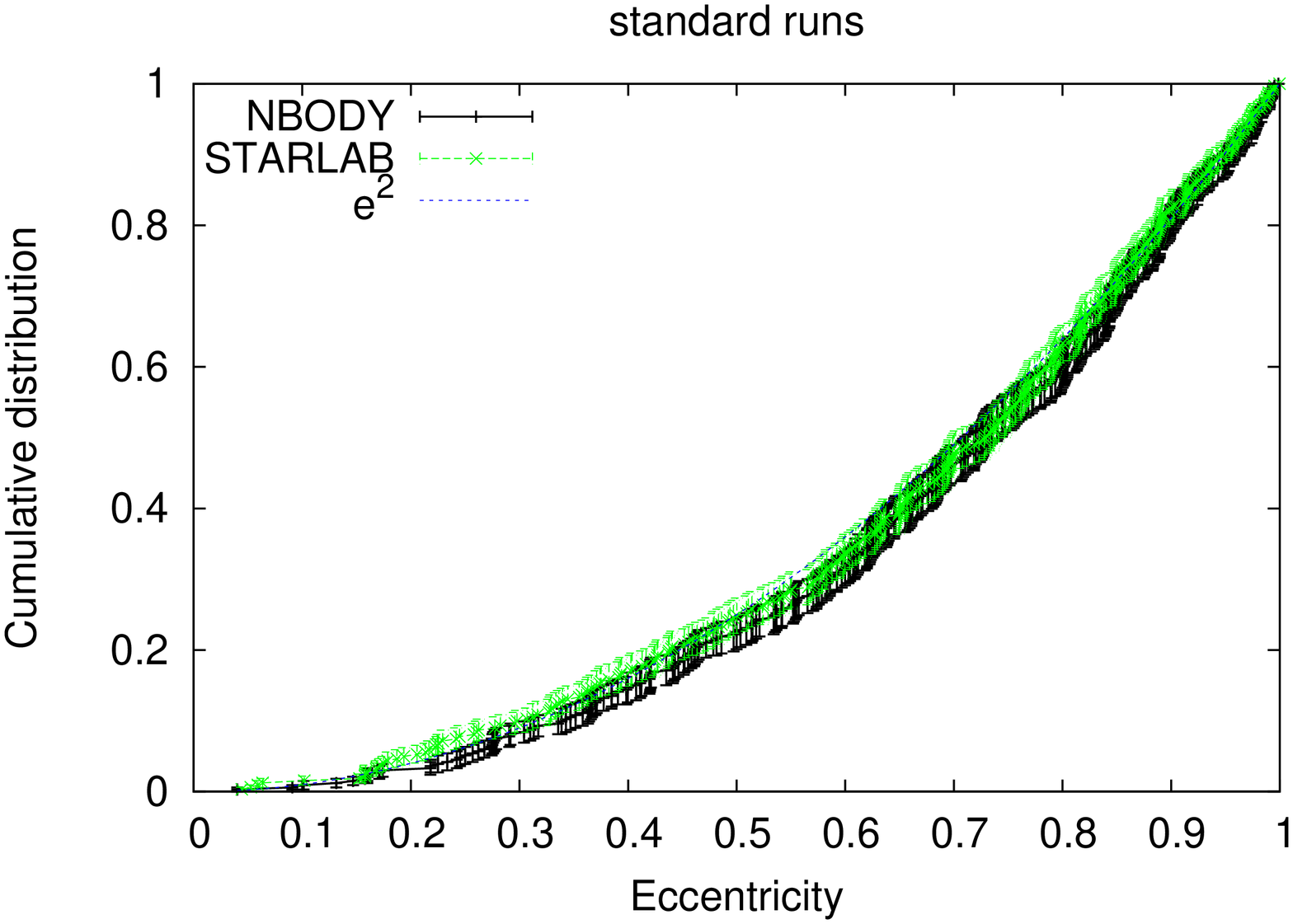} \\  
 		\includegraphics[angle=270,width=0.4\linewidth]{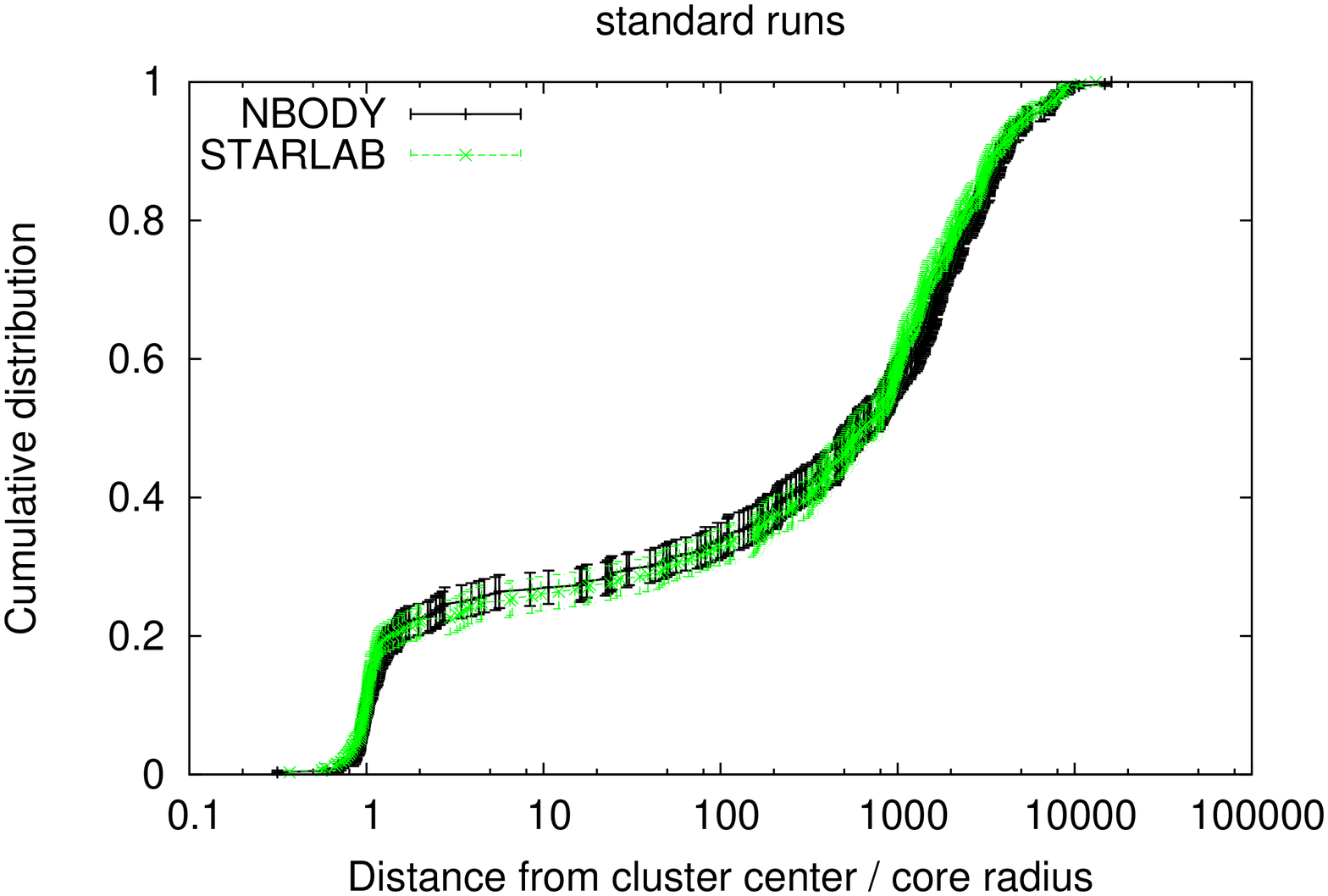} &
 		\includegraphics[angle=270,width=0.4\linewidth]{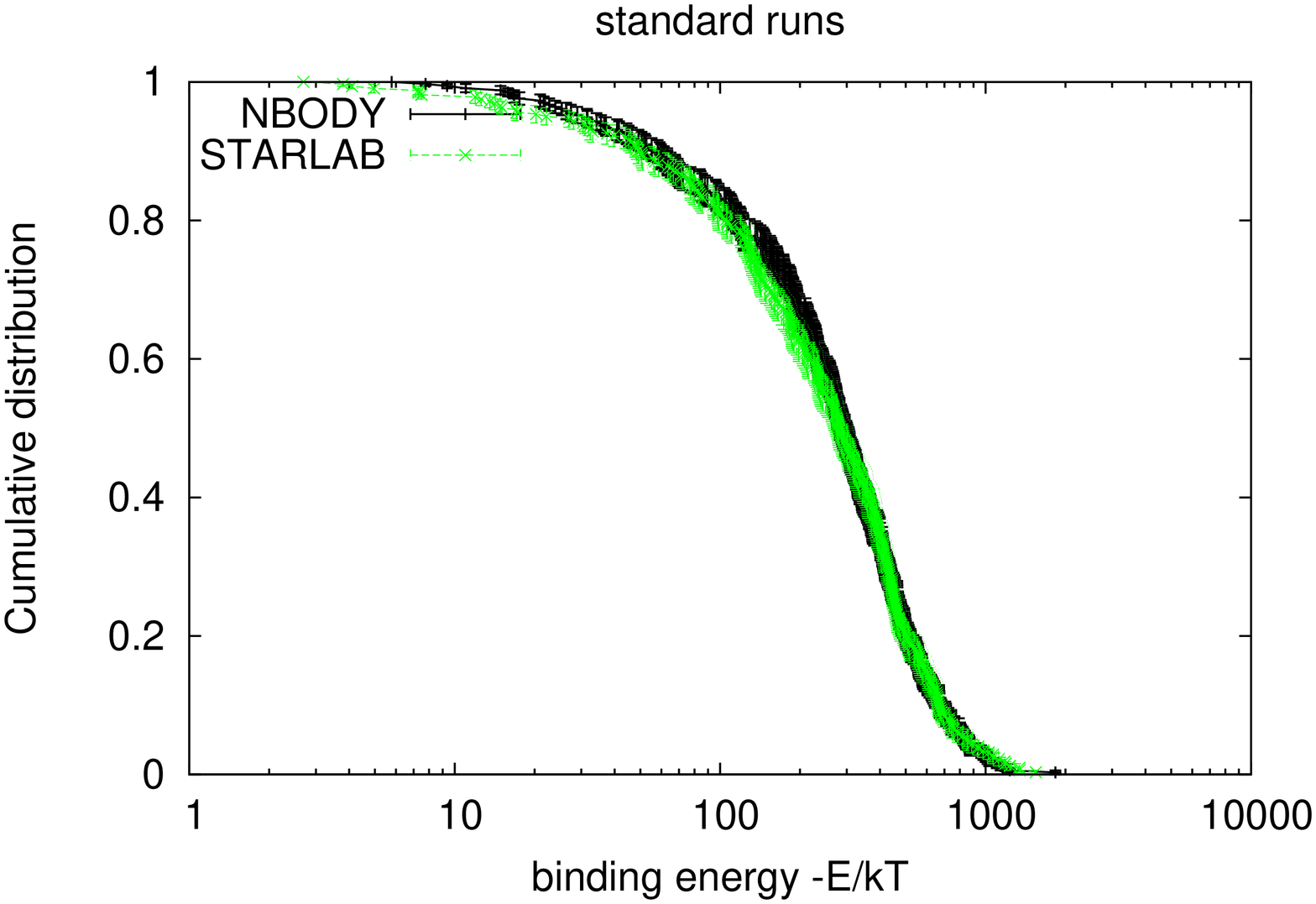} \\
	\end{tabular}

	\caption{Comparison of binary parameters from ``standard runs'' after
	1000 N-body time units (=well after core collapse) using  {\sc
	starlab} (green/grey) vs {\sc nbody4} (black). Shown are the
	cumulative distributions of the semi-major axis  (top left), the
	eccentricity (top right), the distance from the cluster centre (in
	units of the cluster's  core radius; bottom left) and the binding
	energy (bottom right). The lines show the data, the error bars give
	the uncertainty ranges from bootstrapping. For the eccentricity, the
	prediction for a thermal distribution (e$^2$) is overplotted.}

	\label{fig:std_bin_par}
\end{center}
\end{figure*}

Fig. \ref{fig:std_struct_par} (top left) shows that
core-collapse occurs at around 320 N-body time units. This coincides
well with the often-used criterion of the first occurrence of a binary
with binding energy higher than 100 kT (see Table \ref{tab:cc_par}). At
the same time, other structural parameters start to change as well (e.g.
the cluster's half-mass radius, shown in Fig. \ref{fig:std_struct_par},
top right), as the occurrence of hard binaries starts heating
the cluster core, propagating outwards, resulting in an irreversible
overall expansion of these rather low-mass clusters.

At the time of core collapse, also both kinetic and potential energy of
the cluster as a whole change their behaviour: the potential energy
increases, while the kinetic energy decreases slightly, as energy gets
increasingly locked up in binaries (Fig. \ref{fig:std_struct_par}, lower
panels). This effect will be studied further in Sect.
\ref{sec:reswobin}, where the effects of rebuilding the binary tree
structure is discussed.

The results obtained with {\sc starlab} and with {\sc nbody4} lie for
most studied parameter within the combined uncertainty ranges. However,
for some parameters small systematic offsets seem to be present. To test
their significance we developed a bootstrapping algorithm for comparing
functions, described in Sect. \ref{sec:bootstrap}.

The results from this bootstrapping test are presented in Table
\ref{tab:boot_results}. In this Table we give for various parameters the
fraction of test runs made with the same settings (i.e. representing the
null hypothesis of a unique parent distribution) that are more deviating
(i.e. which have larger $\Delta_{12}$ respectively $\Gamma_{12}$) than
the results for this parameter from the main simulations using {\sc
nbody4} or {\sc starlab}.

Except for the total energy E$_{\rm tot}$ (and the conservation of
the total energy $\delta$E$_{\rm tot}$), which will be discussed below
(Sect. \ref{sec:resEC}), solely the core radius evolutions (and
quantities calculated from the core radius) are significantly
discrepant. This discrepancy originates from a kink in the median
temporal core radius evolution for the {\sc nbody4} simulations. This
kink could not be traced back to any kink/jump in individual runs (on
the contrary, some {\sc starlab} runs have stronger jumps than any of
the {\sc nbody4} runs). We rather expect this kink to be an unfortunate
cumulative stochastic effect. This is supported by the fact that after
the kink the temporal dependencies from {\sc starlab} and {\sc nbody4}
continue to evolve in parallel, though offset by the amount the kink
caused. The differences in the evolutions of the kinetic energies appear
to be larger than for the core radii, however, as also the uncertainties
are larger these differences are not statistically significant.

In Fig. \ref{fig:std_bin_par} we study the distribution of the
parameters describing the dynamically created binaries. For the 1k
standard runs core collapse occurs at approximately 320 N-body time
units (see Fig. \ref{fig:std_struct_par} and Table \ref{tab:cc_par}). We
show the parameter distributions of binaries present at 1000 N-body time
units, hence well after core collapse. The shown error bars are
estimated from 10000 bootstrap realisations of each dataset. Visually,
the distributions compare well.

In order to quantify differences between the parameter distributions
using either {\sc starlab} or {\sc nbody4} we used a Kuiper test
(\citealt{kuipertest}, i.e. an advanced KS test, for KS test see e.g.
Numerical Recipes \citealt{1992nrfa.book.....P}). The results are
presented in Table \ref{tab:kstest_bin}. Given are the total numbers of
binaries per set of simulations and the Kuiper test results in \%. A
Kuiper test result is the probability that 2 distributions are drawn
from the same parent distribution. Some of the results seem to be
inconsistent with being drawn from the same parent distribution.
However, we have performed a large number of comparisons with the Kuiper
test. This unavoidably leads to the problem of multiple testing, which
can basically be understood such that if we perform 100 independent
tests, a fraction of the order of $10\%$ of p-values below $0.1$ will
arise by chance even if none of the null hypothesis of the tests would
be wrong. Moreover, the small p-values occur for tests with sample sizes
$\approx 50$, which is just at the lower limit for reliable results with
the Kuiper test. Hence, in view of the small number of "significantly
small" p-values (these are marked coloured in Table
\ref{tab:kstest_bin}), {\sl we conclude that we do not find significant
evidence for deviations of interesting size of the properties of
dynamically created binaries from {\sc nbody4} and {\sc starlab}
simulations.}

We also tested the eccentricity  distributions against the common
assumption of a thermal distribution, which is an eccentricity
distribution $\sim$ 2*e, or a  cumulative distribution $\sim$ e$^2$. A
thermal distribution is generally expected based on phase space
arguments (see e.g. \citealt{1975MNRAS.173..729H}). The results are
given in the last two columns of Table \ref{tab:kstest_bin} for {\sc
starlab} and {\sc nbody4} respectively, and show good agreement with a
thermal distribution. However, cumulative  distributions of higher
polynomial  order than e$^2$ are not rejected either by the Kuiper
test.  We therefore used the binary data obtained as by-product from the
calculations of the bootstrap test clusters, using {\sc starlab} only.
The number of runs is a factor 6 higher than for our main simulations,
hence statistics also for the binaries is greatly enhanced (total number
of binaries is 2018, compared to 323 for the main simulations). We test
their cumulative eccentricity distributions against a number of
power-law distributions e$^\alpha$. For the STD test clusters we find a
range in $\alpha$ = 2.1 -- 2.8 with Kuiper test probabilities $>$ 10\%,
with a probability $>$ 95\% for $\alpha$ = 2.3 -- 2.5, hence
significantly biased towards larger eccentricities than a thermal
distribution would predict (a thermal distribution has a Kuiper test
probability = 4.16\%, hence is significantly rejected). 

We split the whole sample in thirds, based on the semi-major axis, the
distance from the cluster centre and the binding energy. However, due to
the reduction of the number of binaries in each of these subsets, the
Kuiper test does not reject the null hypothesis of the eccentricity
distributions being thermal on a significant level (except for the
subset of binaries with intermediate semi-major axes, which has a
p-value of 2.5\%). The p-value curves are too broad to derive any
trends.

The general agreement between the data obtained using either {\sc
starlab} or {\sc nbody4} is good.

The distributions just after core collapse give comparable results
(except for the  spatial distribution, as the binaries did not yet have
enough time to escape the cluster centre significantly), although the
number of binaries is smaller (i.e. statistics is poorer).

\subsection{Results for the ``MF10 runs''}
\label{sec:resmf10}

\begin{figure*}
\begin{center}
  \vspace{-0.5cm}
  \hspace{1.2cm}
  \begin{tabular}{cc}	
	\includegraphics[angle=270,width=0.4\linewidth]{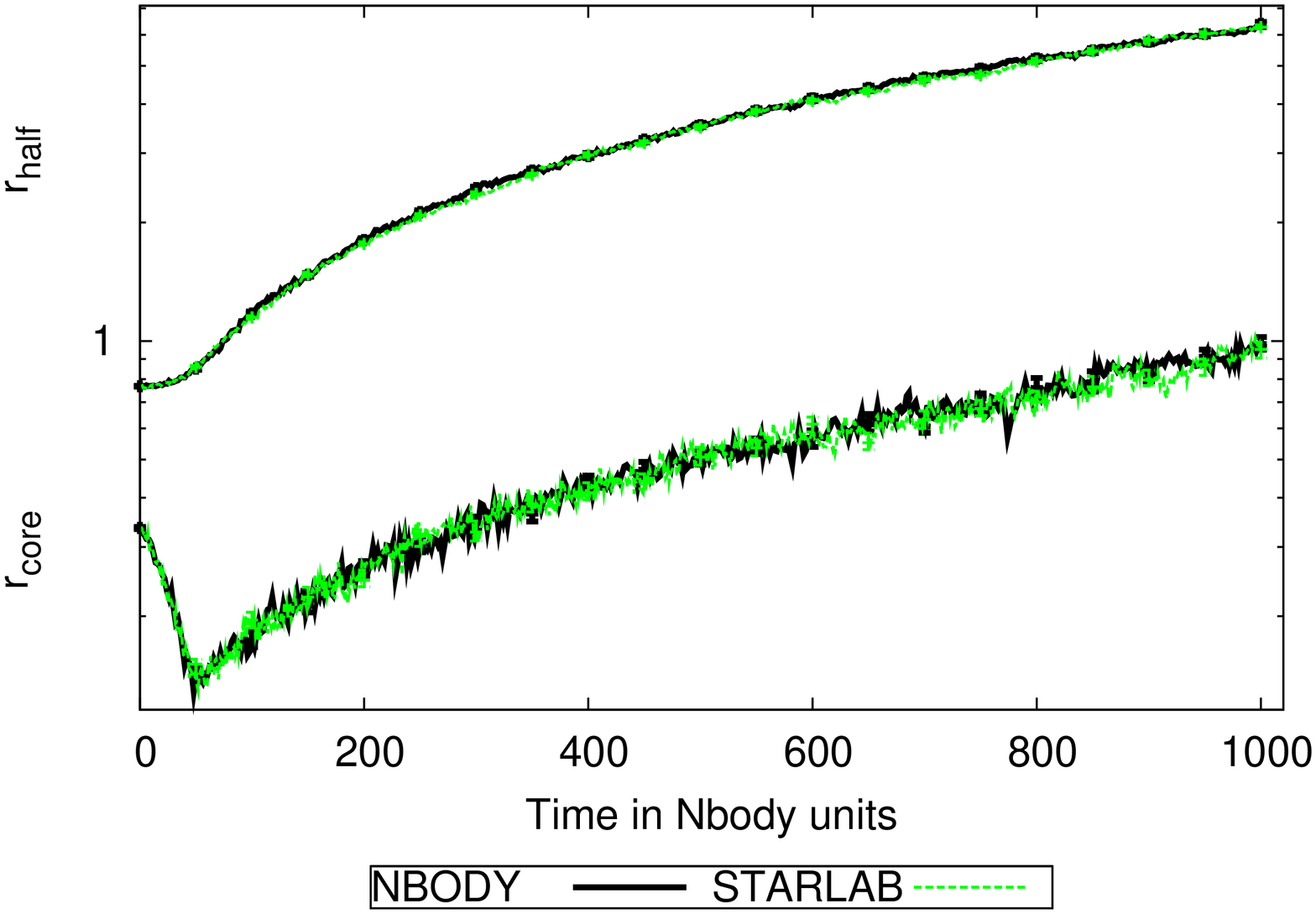} & 
	\includegraphics[angle=270,width=0.4\linewidth]{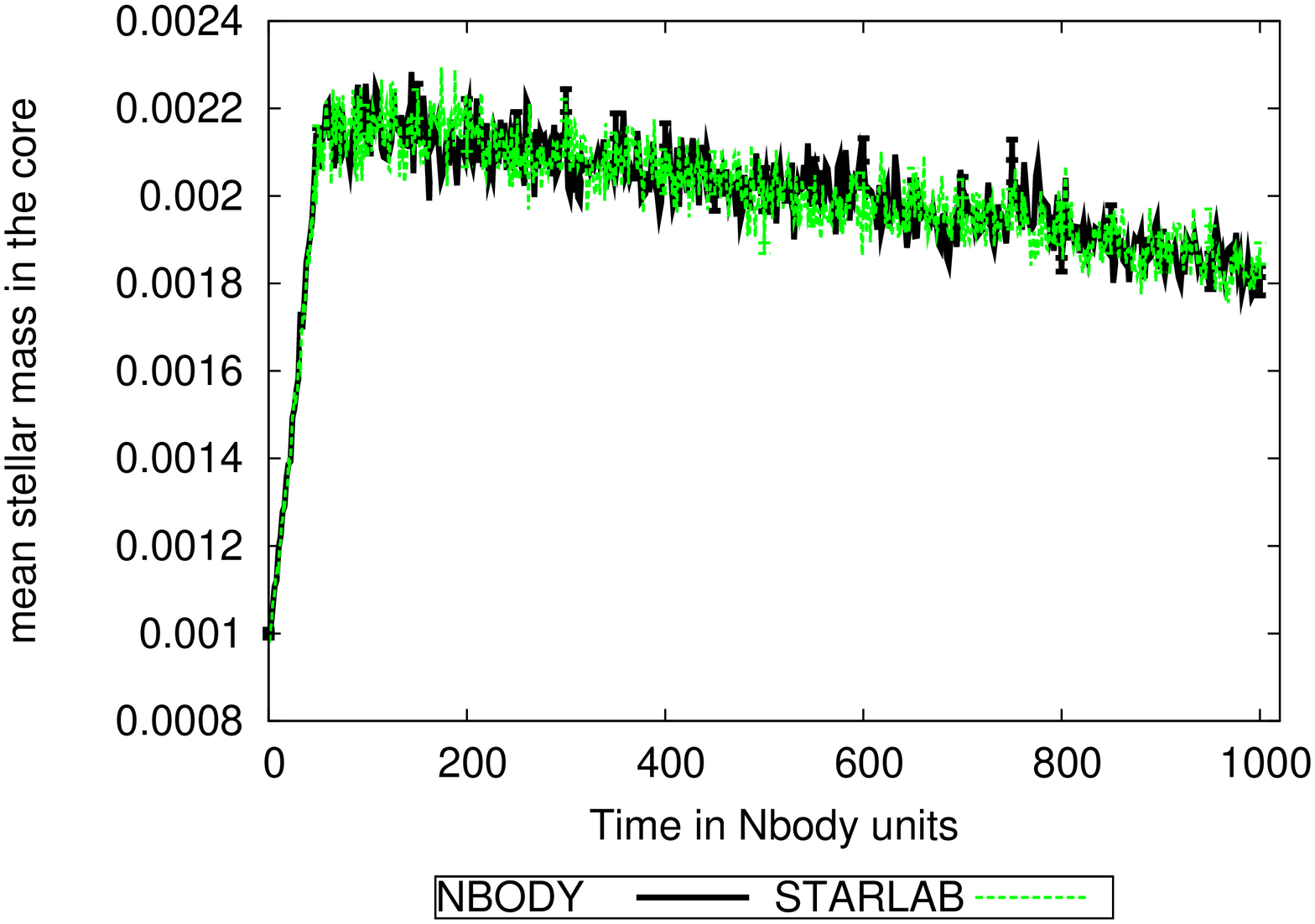} \\
	 \includegraphics[angle=270,width=0.4\linewidth]{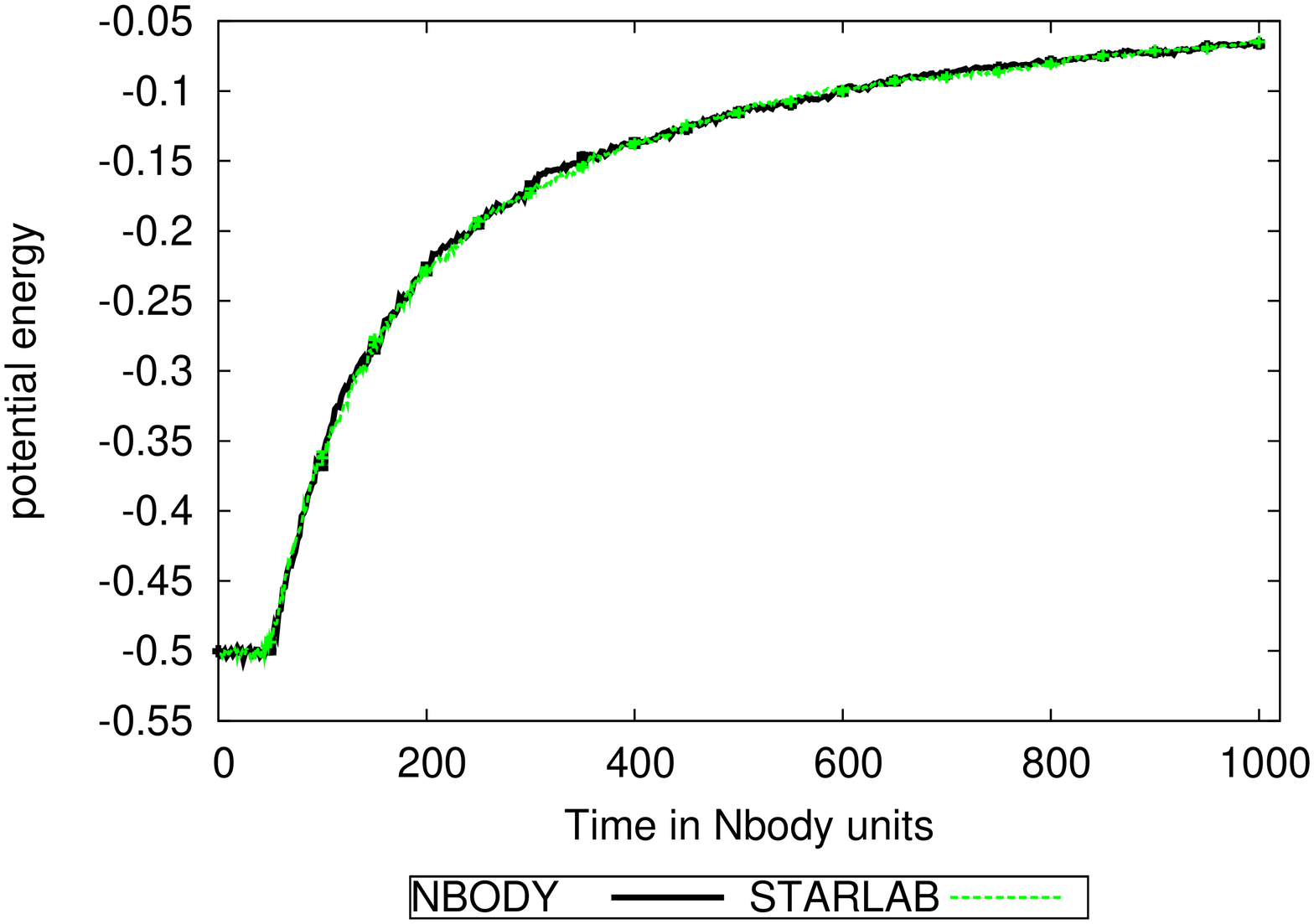} &
	 \includegraphics[angle=270,width=0.4\linewidth]{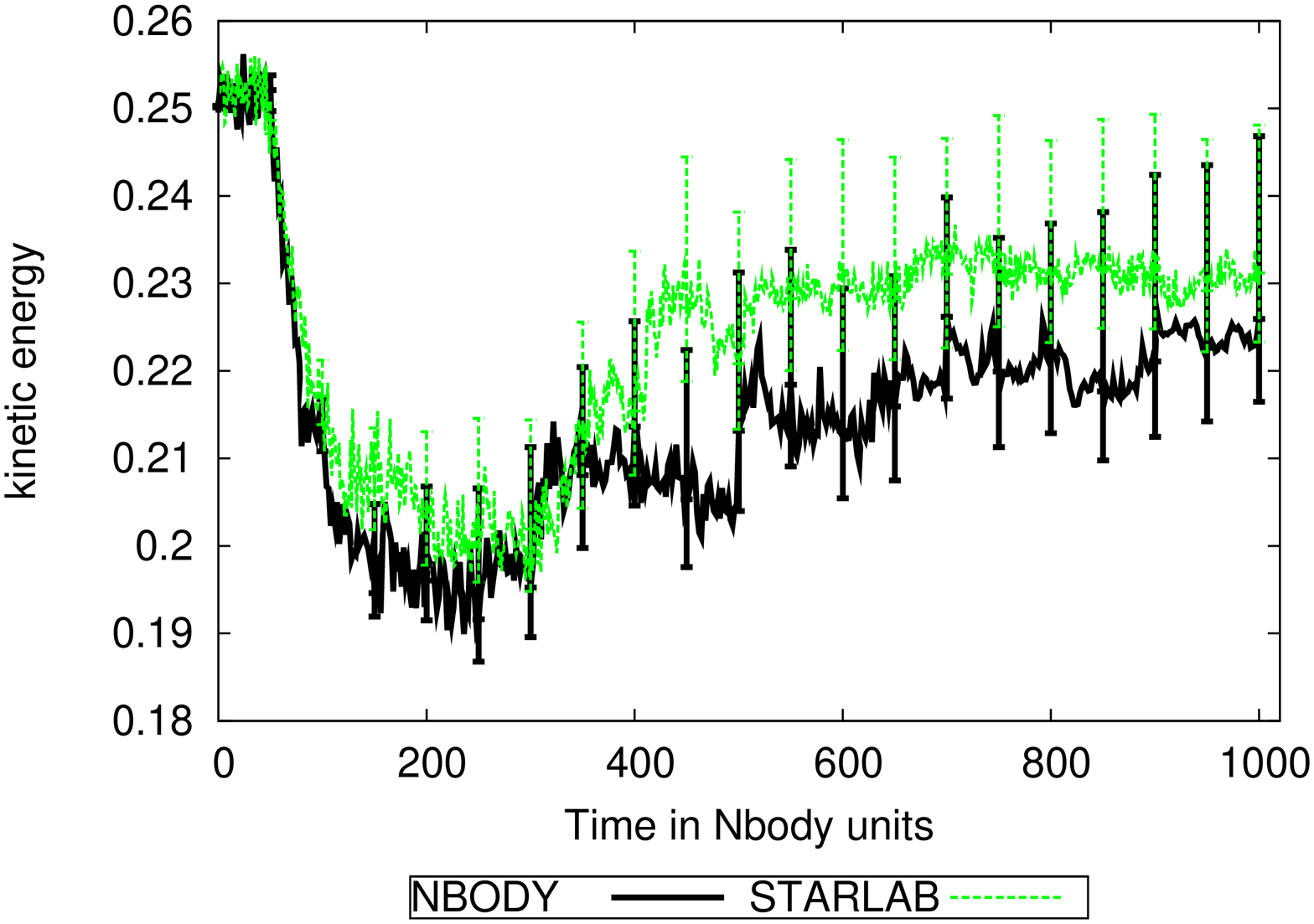} \\  
  \end{tabular}
\end{center}

\caption{Comparison of ``MF10 runs'' simulations using {\sc starlab}
(green/grey) vs {\sc nbody4} (black).  The lines show the median values,
the error bars give the uncertainty ranges from the 50 individual runs.
Shown are the time evolutions of the core radius (top left,
bottom lines), half-mass radius (top left, top lines), the mean
object mass in the core (top right), potential energy (lower
left panel) and kinetic energy (bottom right).}

\label{fig:wmf10_struct_par}

\begin{center}
	\begin{tabular}{cc}
		\includegraphics[angle=270,width=0.4\linewidth]{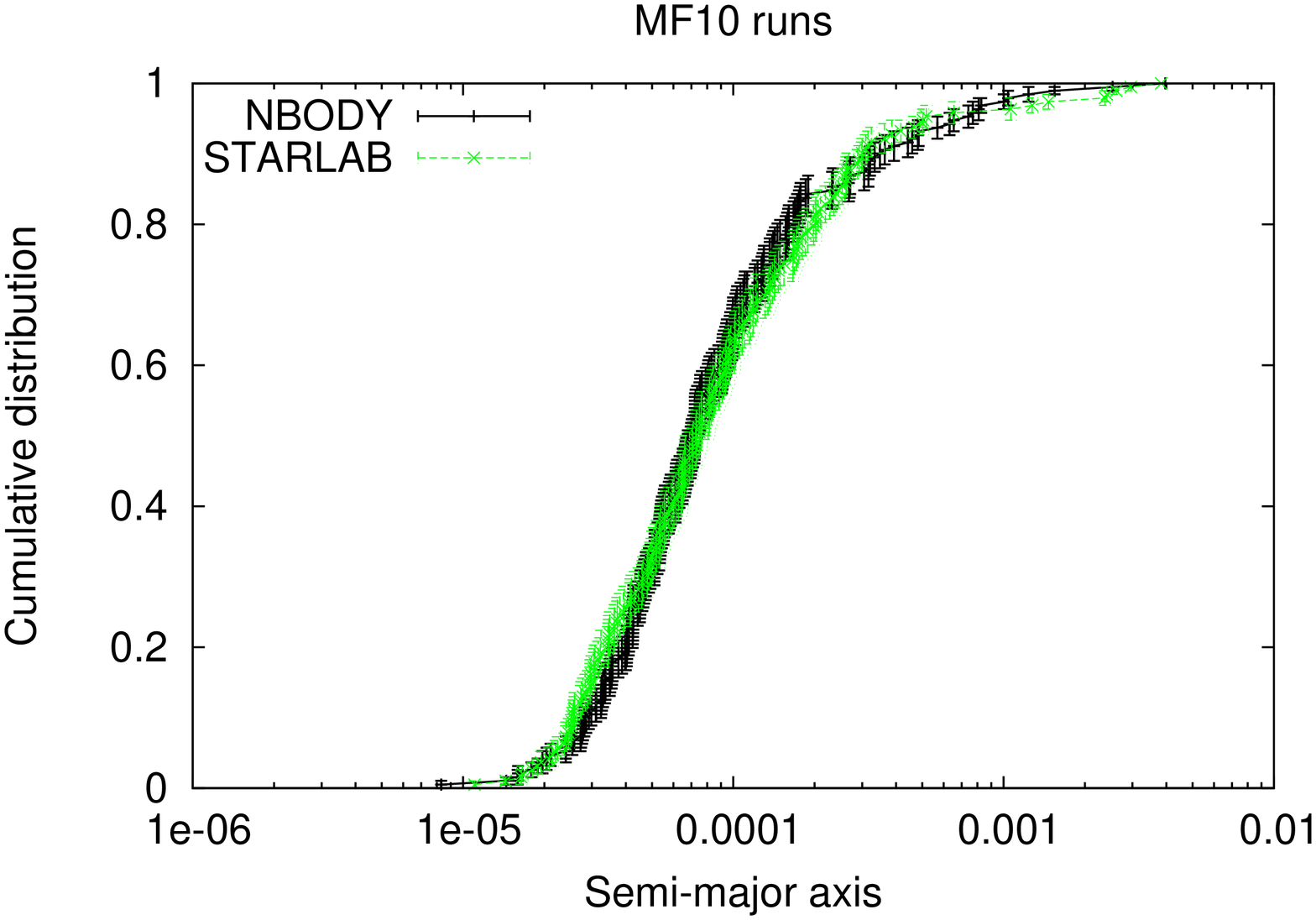} &
 		\includegraphics[angle=270,width=0.4\linewidth]{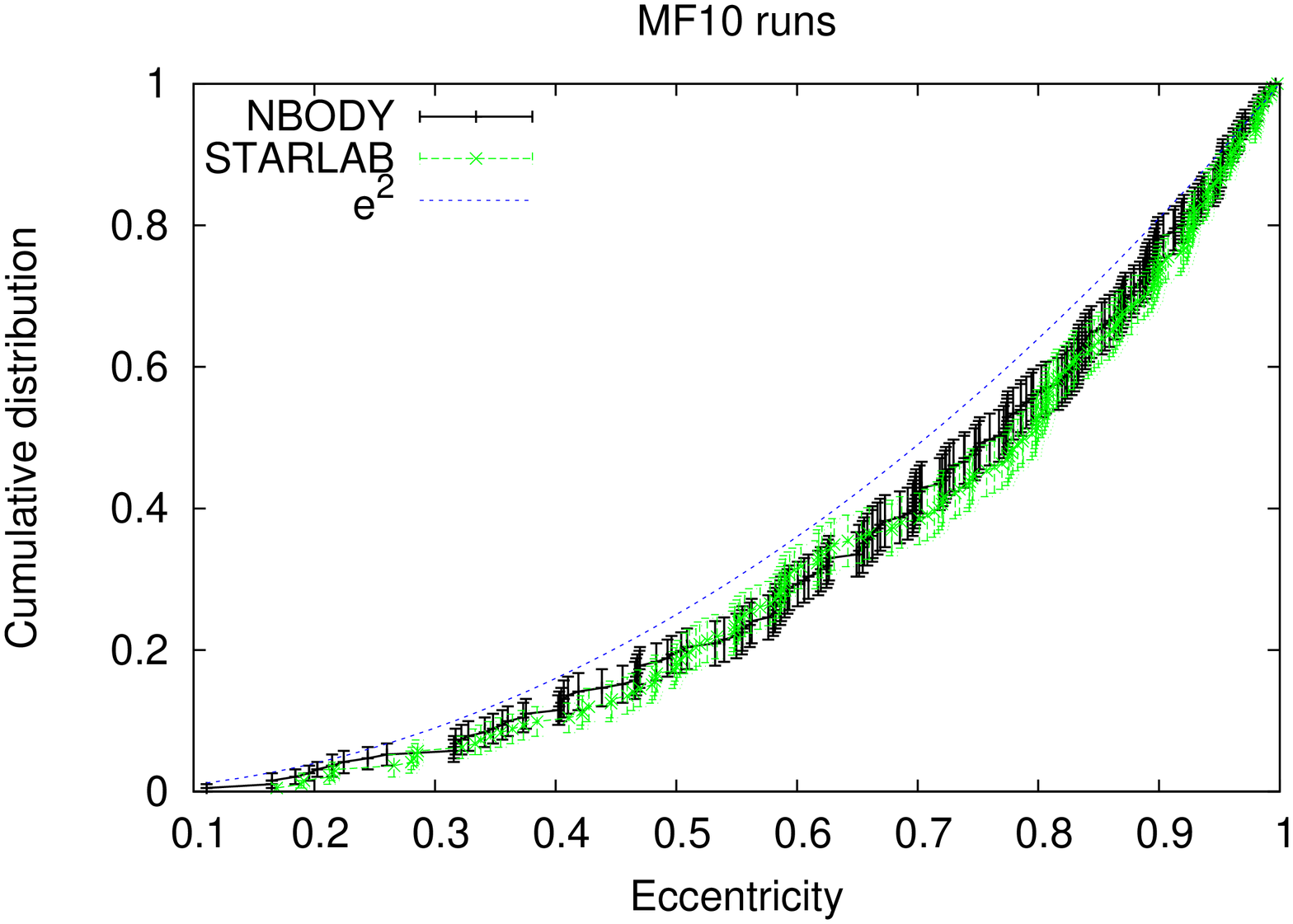} \\  
 		\includegraphics[angle=270,width=0.4\linewidth]{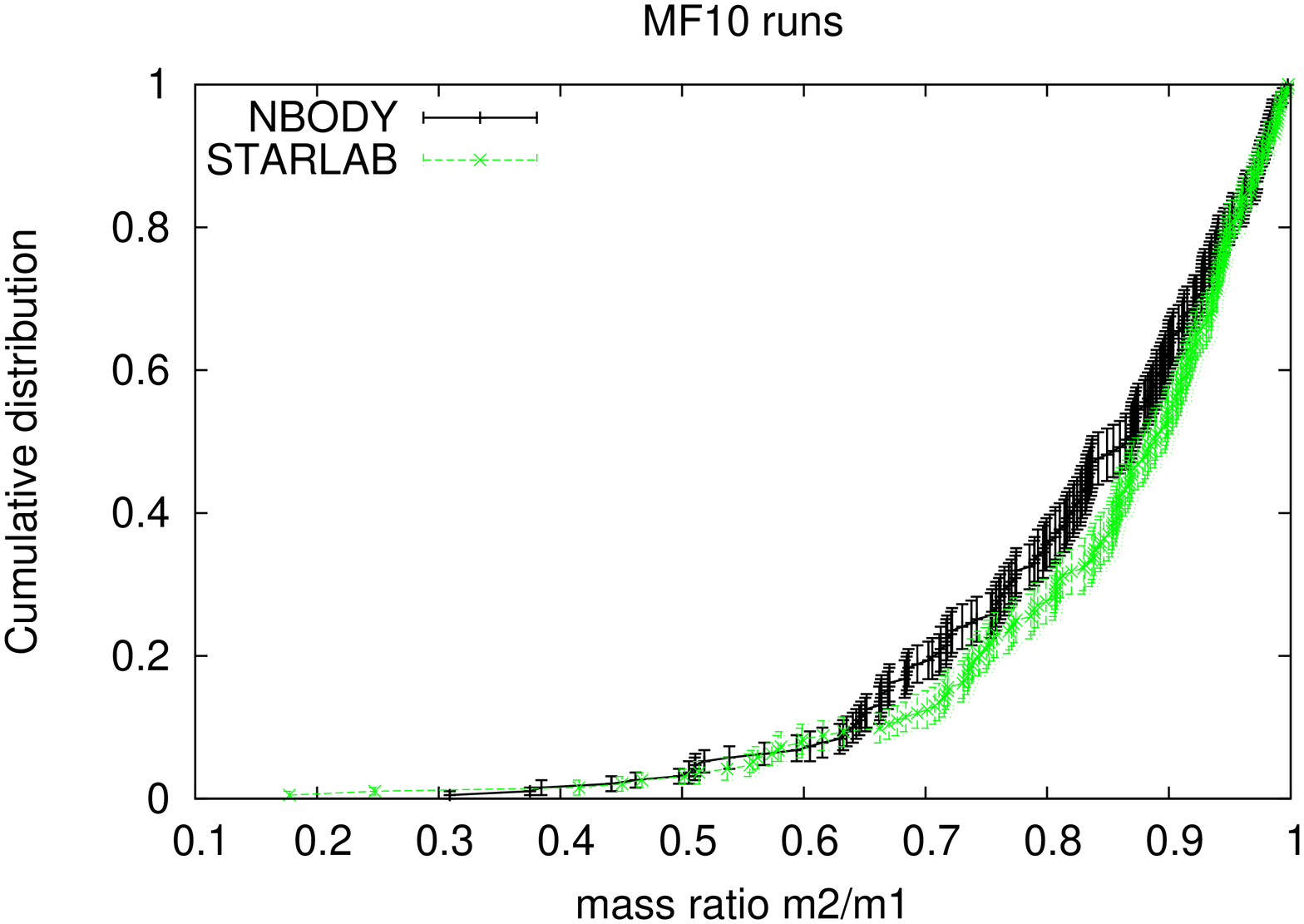} &
 		\includegraphics[angle=270,width=0.4\linewidth]{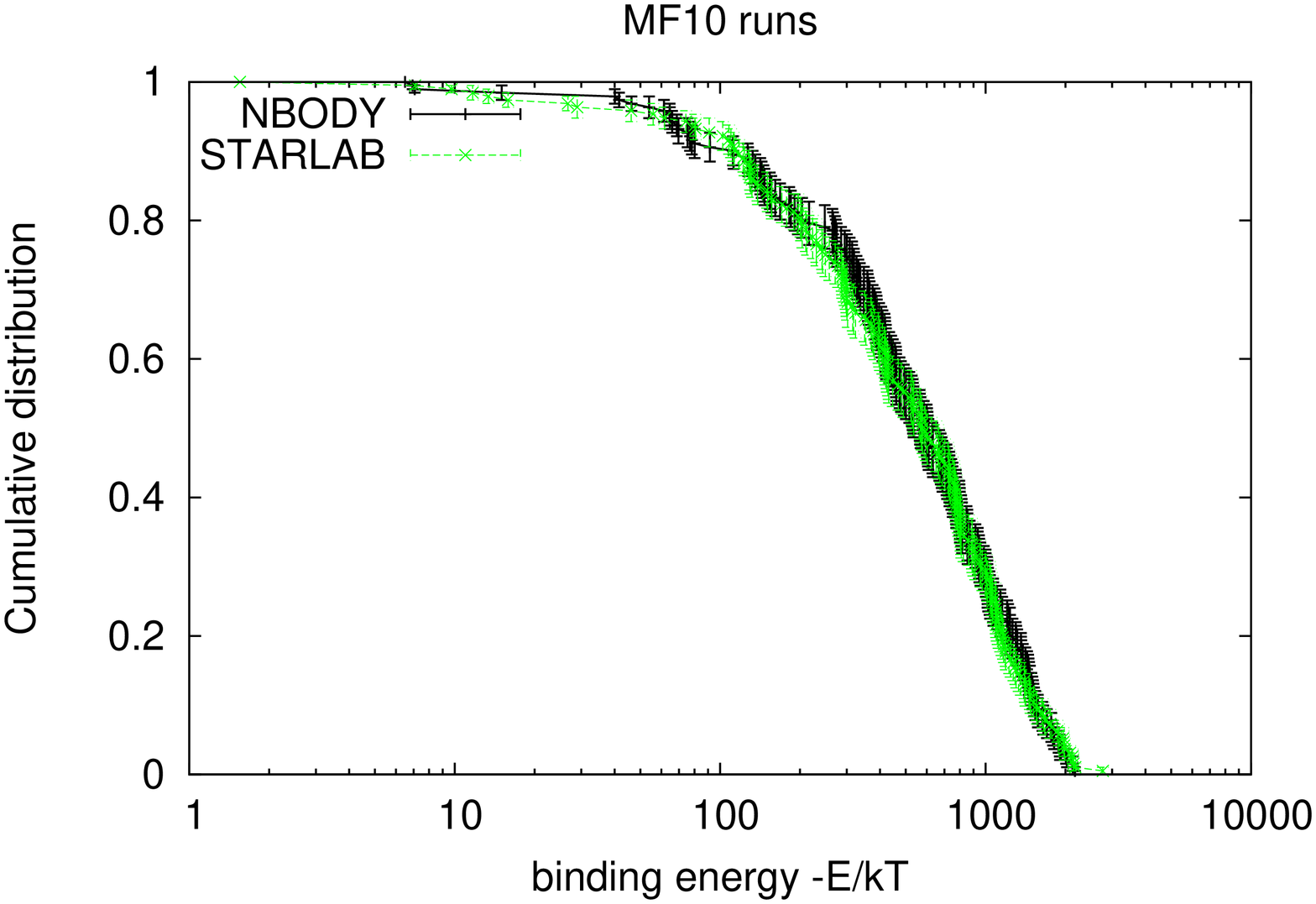} \\
	\end{tabular}

	\caption{Comparison of binary parameters from ``MF10 runs'' after
	1000 N-body time units (=well after core collapse) using  {\sc
	starlab} (green/grey) vs {\sc nbody4} (black). Shown are the
	cumulative distributions of the semi-major axis  (top left), the
	eccentricity (top right), the secondary-to-primary mass ratio (bottom
	left) and the  binding energy (bottom right). The lines show the
	data, the error bars give the uncertainty ranges from bootstrapping.}

	\label{fig:wmf10_bin_par}
\end{center}
\end{figure*}

The results for the ``MF10 runs'' are presented in Fig.
\ref{fig:wmf10_struct_par} - \ref{fig:wmf10_bin_par}.

For these simulations core collapse occurs at around 60-70 N-body time
units (see Fig. \ref{fig:wmf10_struct_par} [upper left panel] and Table
\ref{tab:cc_par}). Qualitatively both the structural behaviour and the
properties of the binaries compare well with the standard case, except
for the speed up the mass function causes. Only the binary binding
energies are higher as compared to the standard runs (by a factor $\sim$
2), although the form of the cumulative distribution of the binding
energies is comparable.

Both the bootstrap test for structural/energy parameters (except for
the  energy conservation, see Sect. \ref{sec:resEC}) and the Kuiper test
for the binary properties prove the very good agreement between the
results obtained from the {\sc nbody4} and {\sc starlab} simulations.

We tested again the hypothesis of a thermal eccentricity distribution of
the dynamically created binaries. For both the {\sc nbody4} and the
{\sc  starlab} main runs, the Kuiper test yields probabilities which do
not reject the hypothesis of a thermal eccentricity distribution. Again,
we used the binary data obtained as by-product from the calculations of
the bootstrap test clusters, using {\sc starlab} only. The total number
of binaries is 1207, compared to 191 for the main simulations. We test
their cumulative eccentricity distributions against a number of
power-law distributions e$^\alpha$. For the MF10 test clusters we find a
range in $\alpha$ = 2.4 -- 3.0 with Kuiper test probabilities $>$ 10\%,
with a maximum probability = 80.7\% for $\alpha$ = 2.8, hence
significantly biased towards larger eccentricities than a thermal
distribution would predict (a thermal distribution has a Kuiper test
probability = 0.06\%, hence is highly significantly rejected). 

We again split the whole sample in thirds, now also based on the primary
mass and the mass ratio. We find that binaries with small semi-major
axis, high binding energy or massive primaries tend to favour
distributions closer to a thermal distribution than binaries with large
semi-major axis, low binding energy or low-mass primaries. This can
qualitatively be understood by assuming those binaries to be the ones
with the most past encounters, hence the higher probability to
thermalise. While the binaries with large semi-major axis, low binding
energy or low-mass primaries are still highly significantly rejected,
binaries with high binding energy are highly significantly rejected,
binaries with high primary mass are weakly significantly rejected and
binaries with small semi-major axis are not rejected to be consistent
with a thermal distribution. The distance from the cluster centre (both
binaries close to the cluster centre and in the far cluster outskirts
are significantly rejected) and the mass ratio of the binary ($\sim$10
per cent, i.e. very weakly significantly rejected) have only small
effects.

\subsection{Results for the ``MF100 runs''}
\label{sec:resmf100}

\begin{figure*}
\begin{center}
  \vspace{-0.5cm}
  \hspace{1.2cm}
  \begin{tabular}{cc}	
    \includegraphics[angle=270,width=0.4\linewidth]{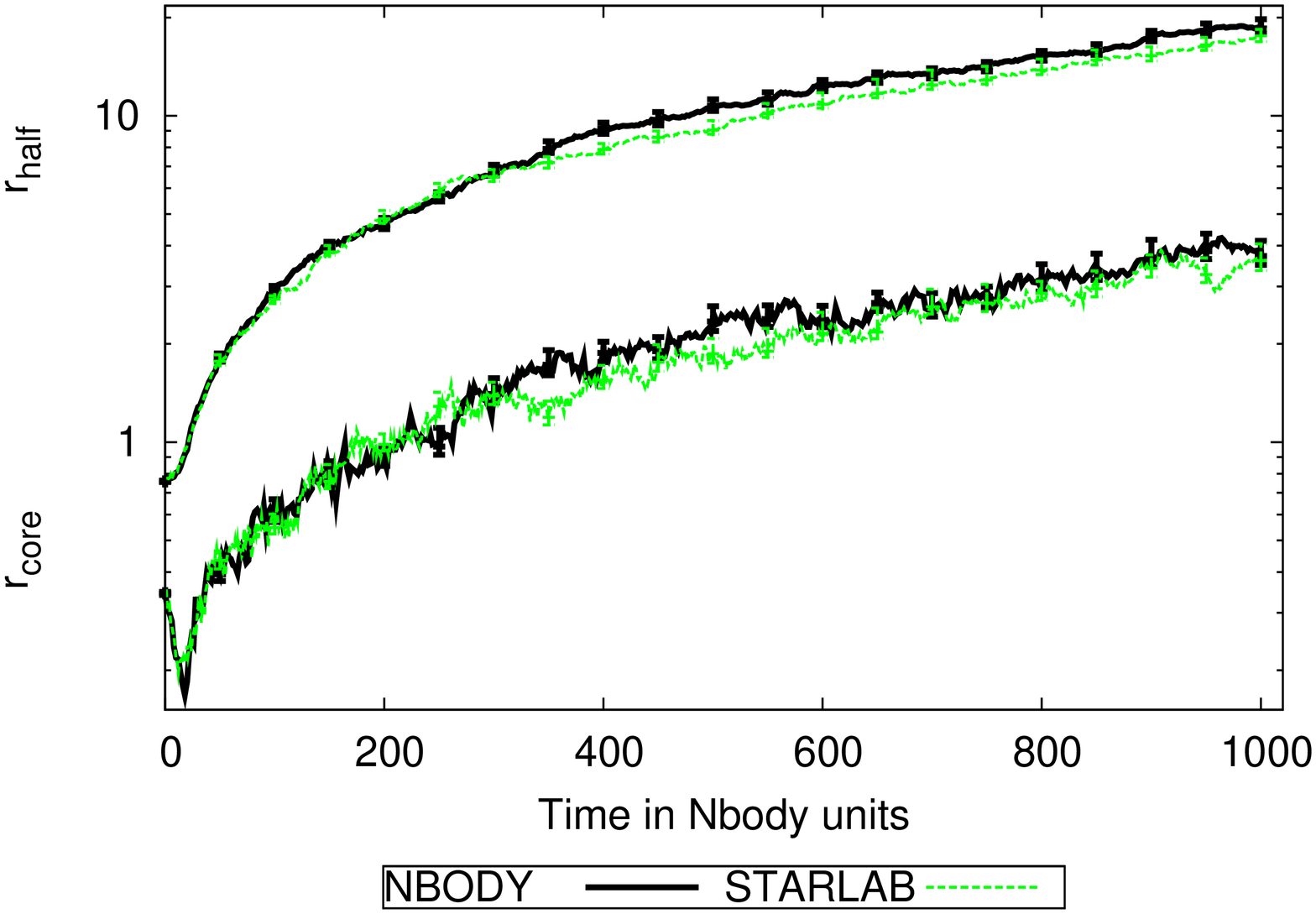} &
	 \includegraphics[angle=270,width=0.4\linewidth]{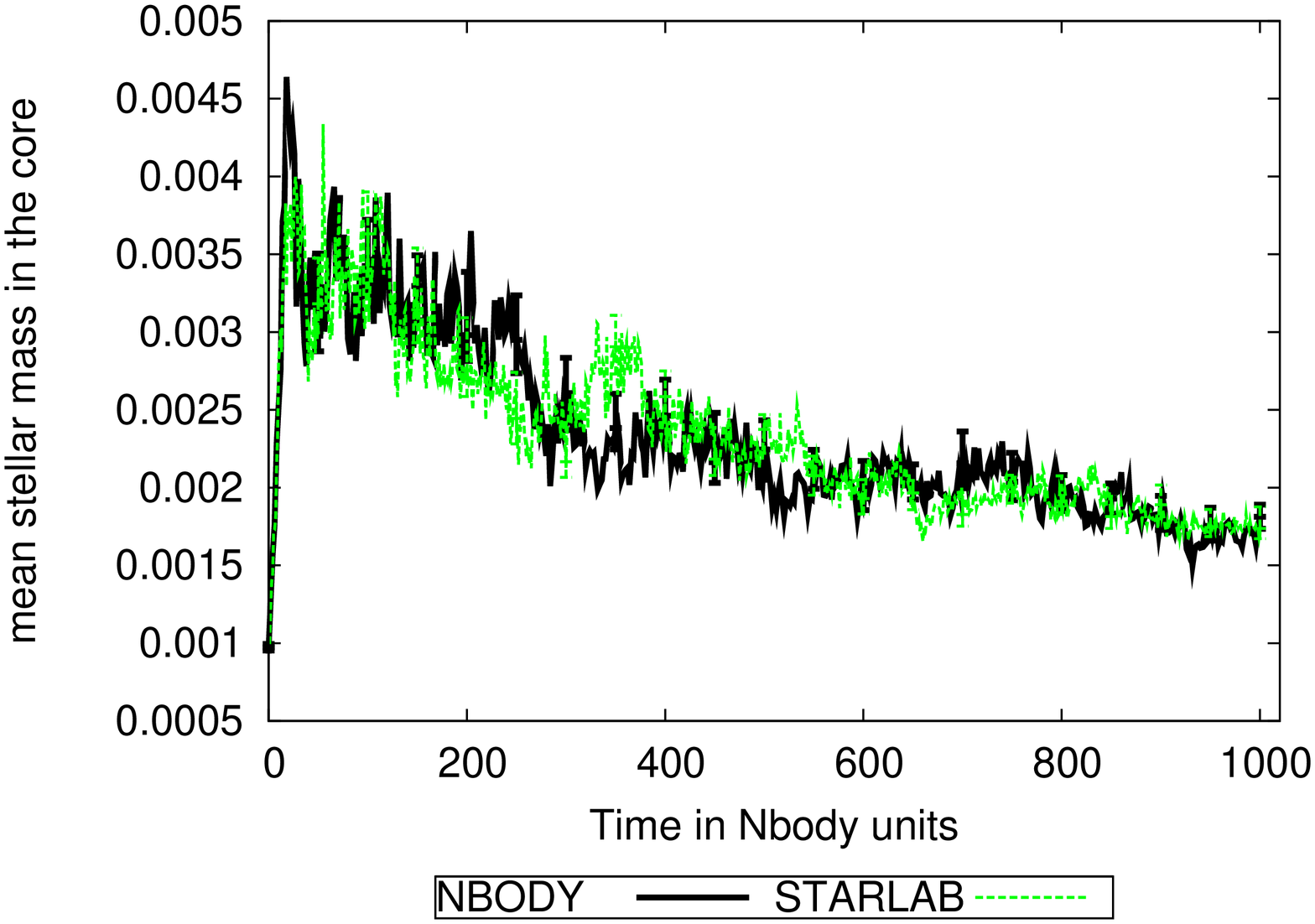} \\
	 \includegraphics[angle=270,width=0.4\linewidth]{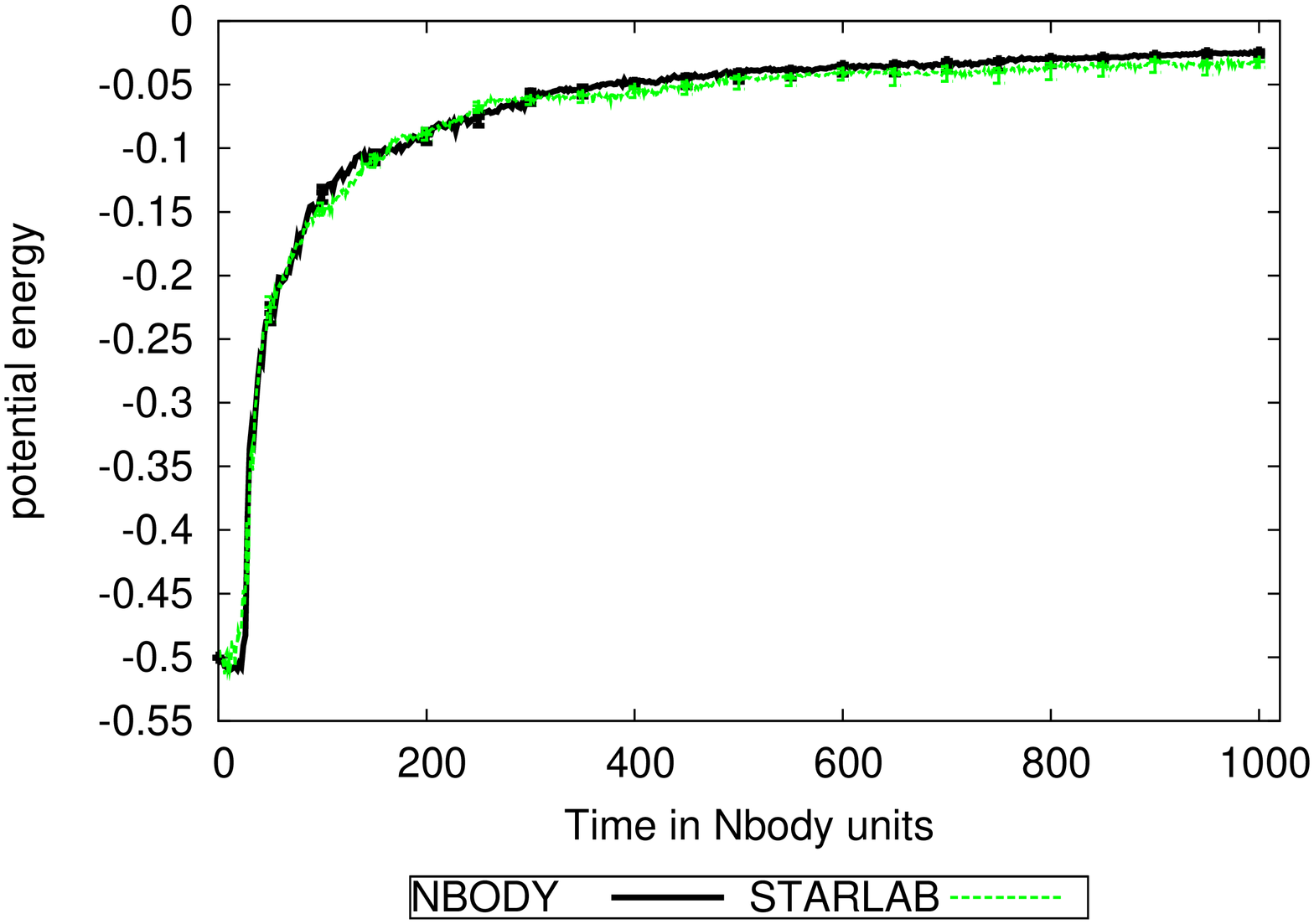} &
	 \includegraphics[angle=270,width=0.4\linewidth]{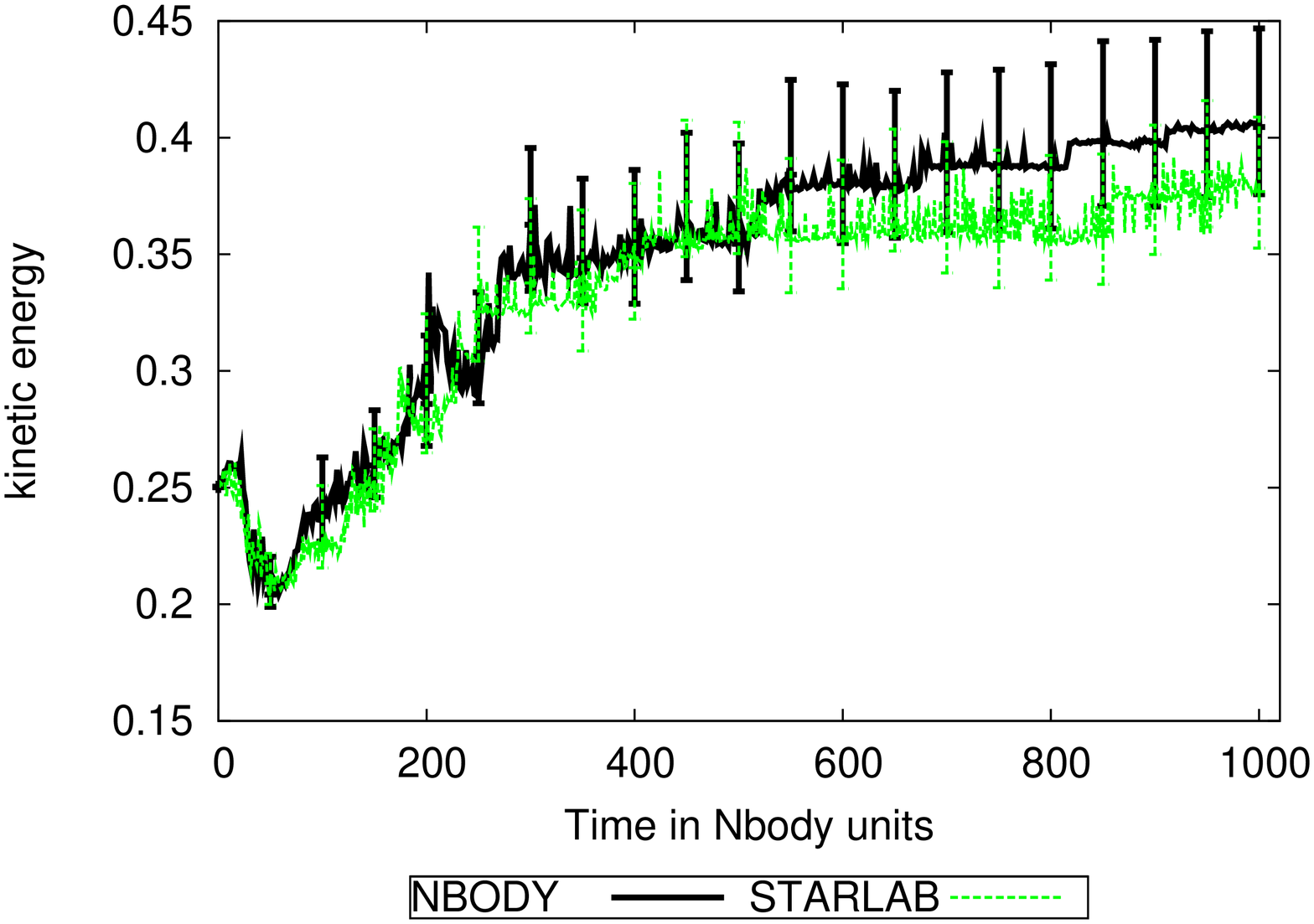} \\  
  \end{tabular}
\end{center}

\caption{Comparison of ``MF100 runs'' simulations using {\sc starlab}
(green/grey) vs {\sc nbody4} (black).  The lines show the median values,
the error bars give the uncertainty ranges from the 50 individual runs.
Shown are the time evolutions of the core radius (top left,
bottom lines), half-mass radius (top left, top lines), the mean
object mass in the core (top right), potential energy (lower
left panel) and kinetic energy (bottom right).}

\label{fig:wmf100_struct_par}

\begin{center}
	\begin{tabular}{cc}
      \includegraphics[angle=270,width=0.4\linewidth]{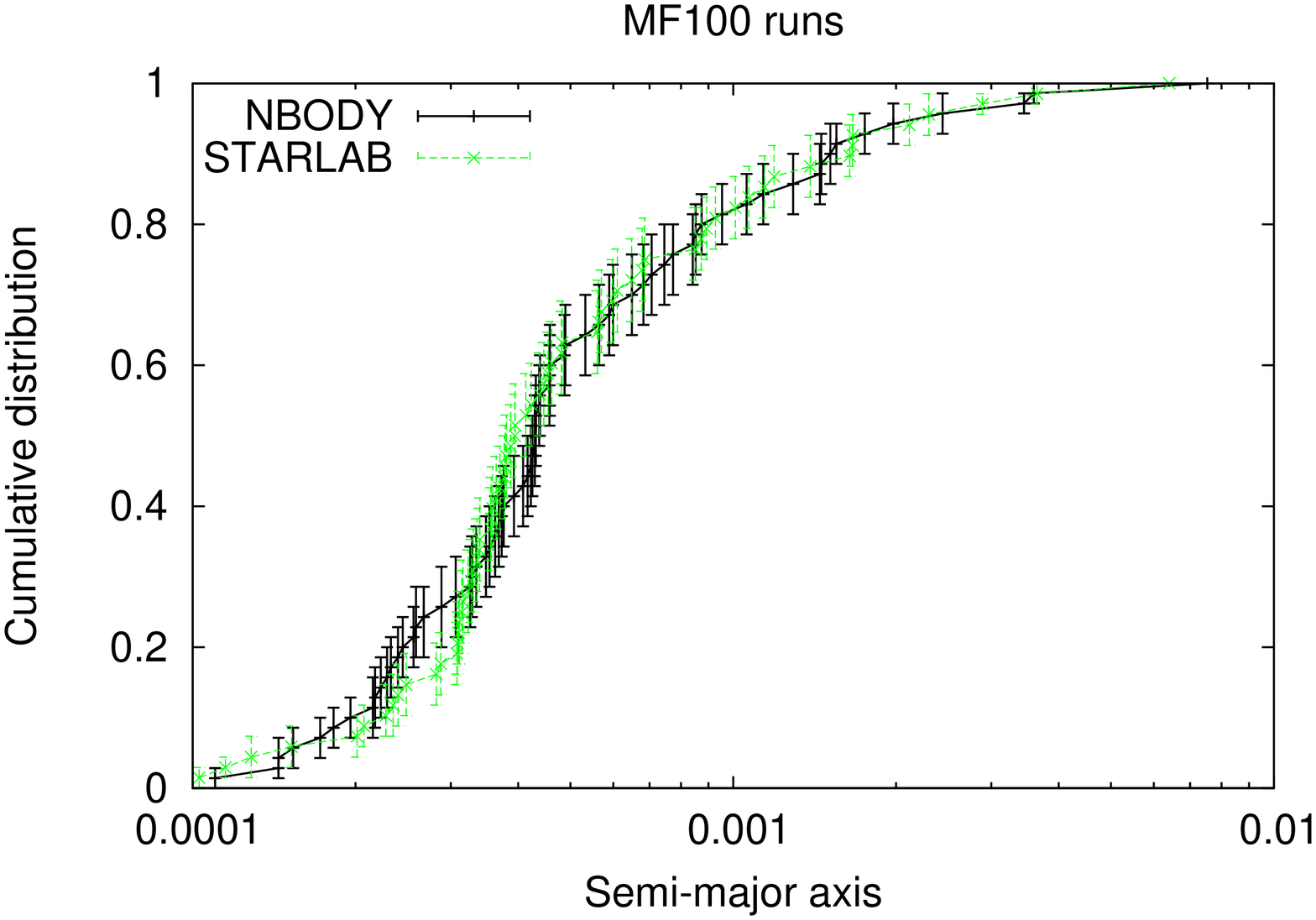} &
 		\includegraphics[angle=270,width=0.4\linewidth]{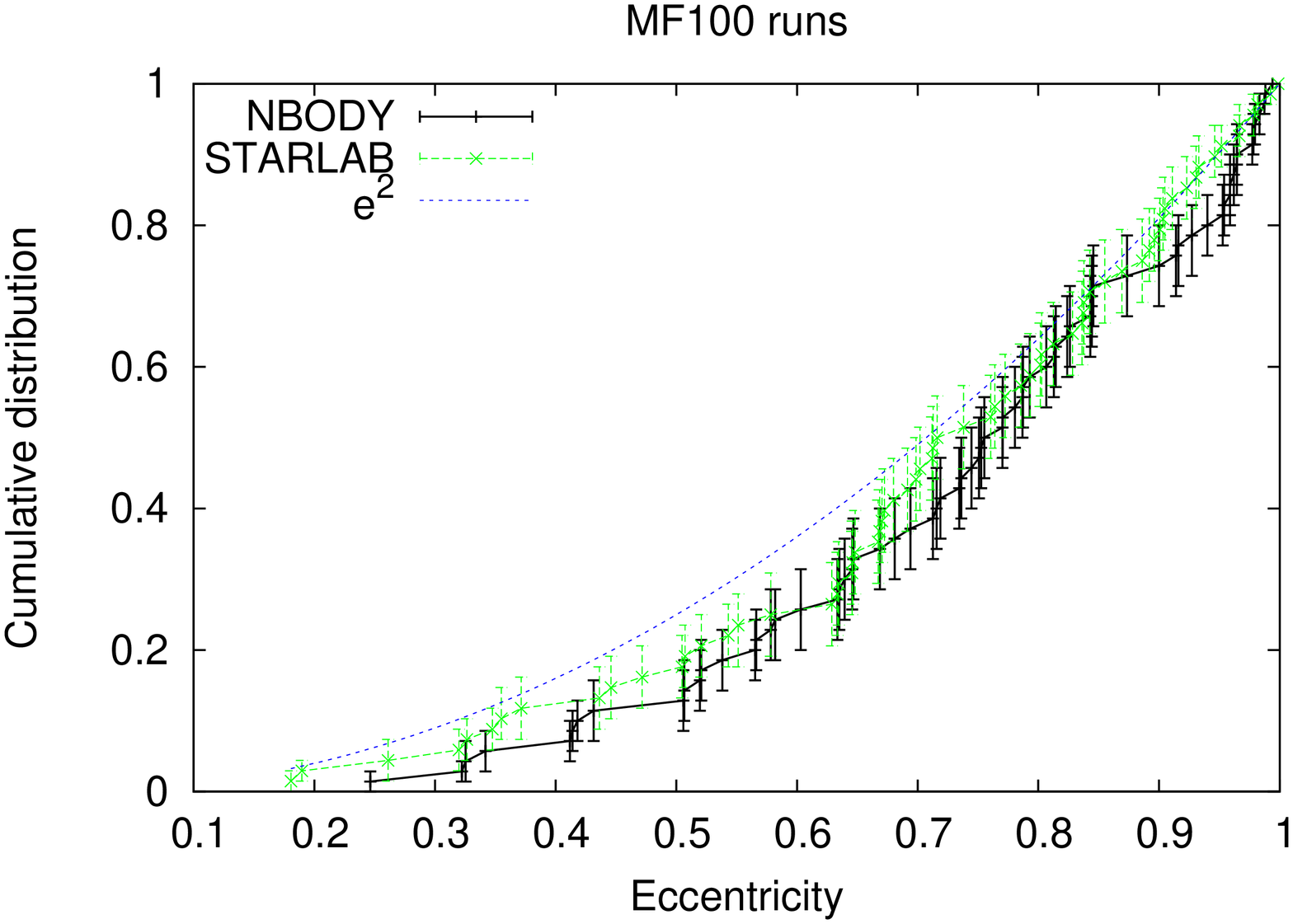} \\  
 		\includegraphics[angle=270,width=0.4\linewidth]{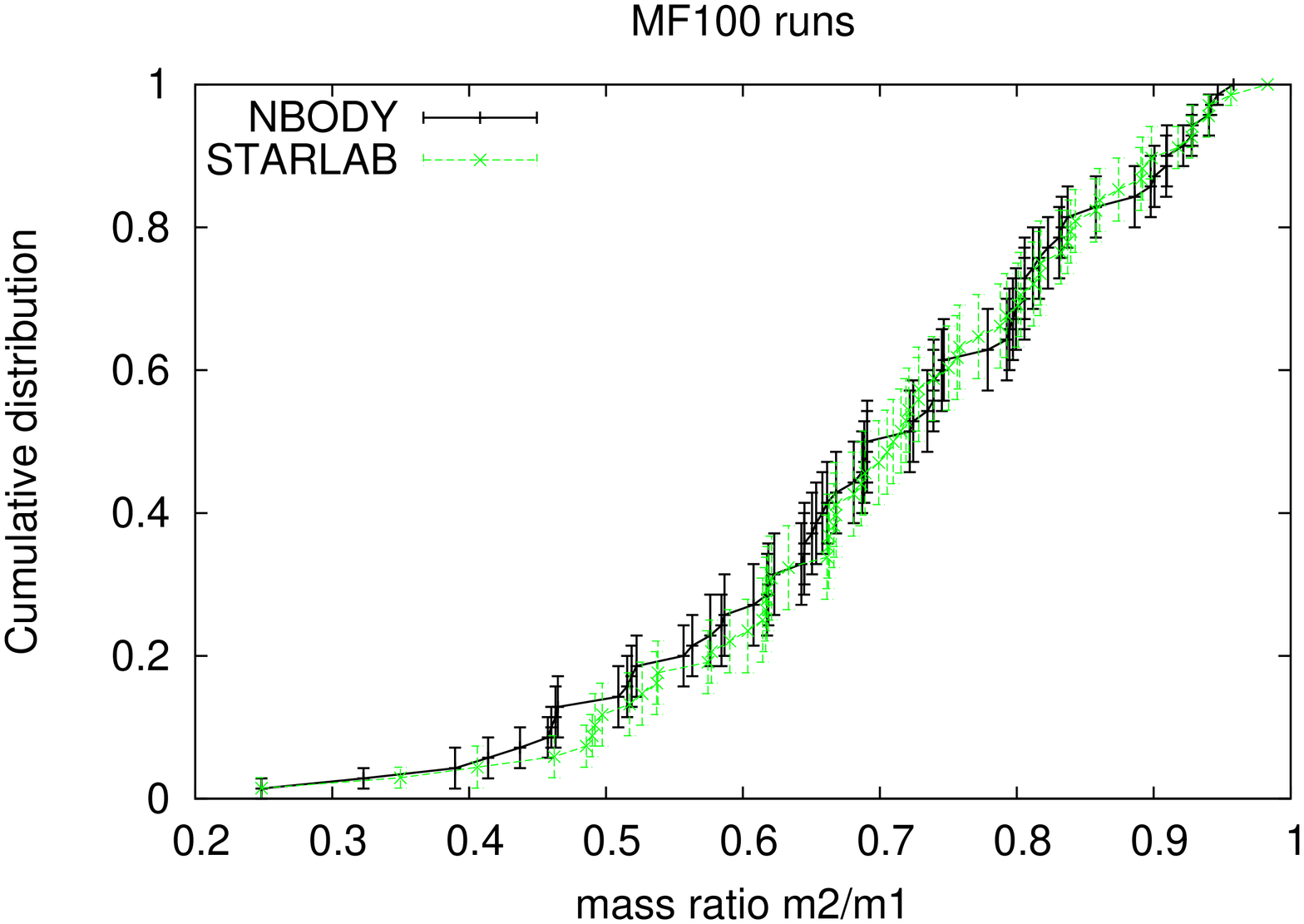} &
 		\includegraphics[angle=270,width=0.4\linewidth]{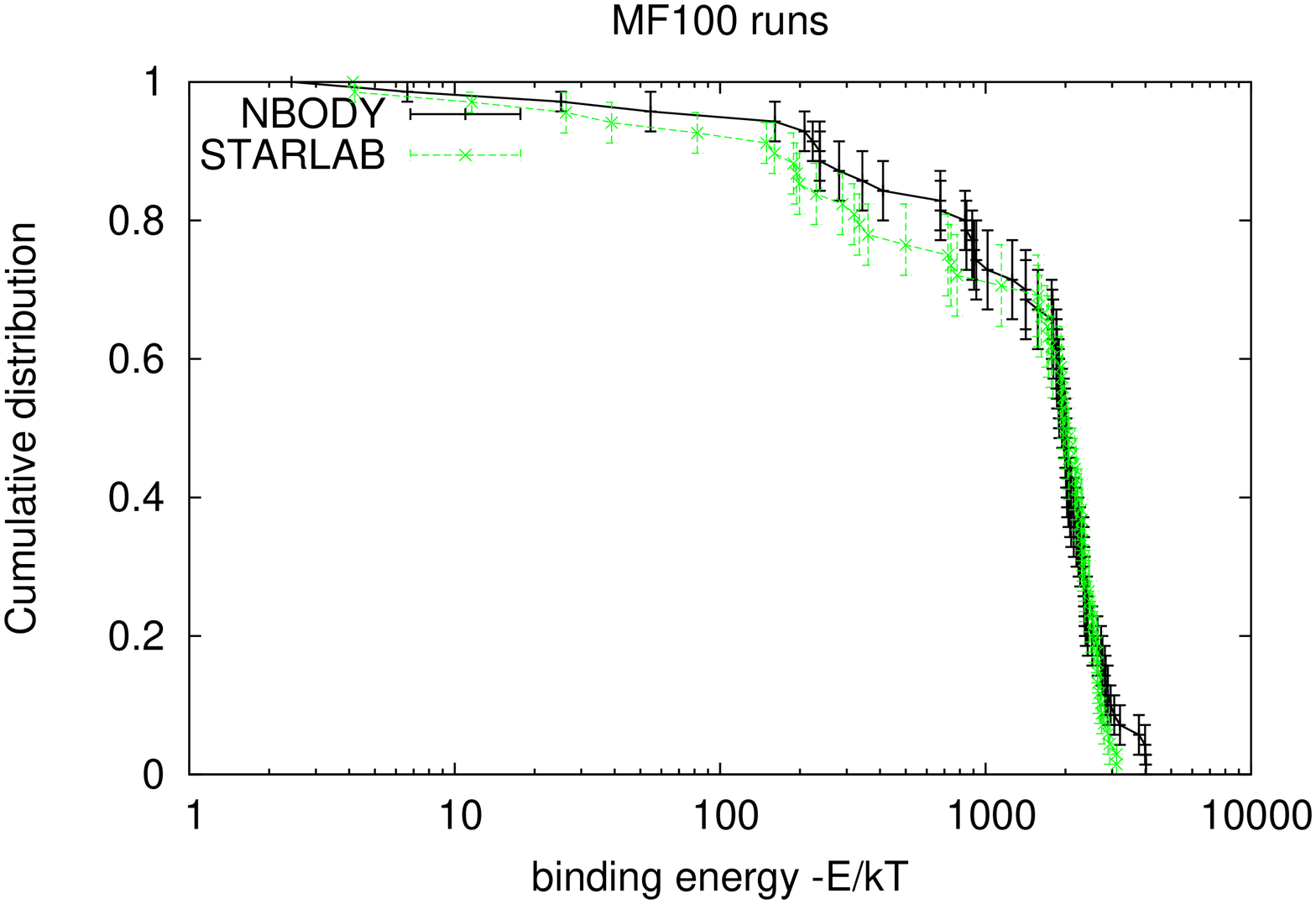} \\
	\end{tabular}

	\caption{Comparison of binary parameters from ``MF100 runs'' after
	1000 N-body time units (=well after core collapse) using  {\sc
	starlab} (green/grey) vs {\sc nbody4} (black). Shown are the
	cumulative distributions of the semi-major axis  (top left), the
	eccentricity (top right), the secondary-to-primary mass ratio (bottom
	left) and the binding energy (bottom right). The lines show the data,
	the error bars give the uncertainty ranges from bootstrapping.}

	\label{fig:wmf100_bin_par}
\end{center}
\end{figure*}

The results for the ``MF100 runs'' are presented in Fig.
\ref{fig:wmf100_struct_par} - \ref{fig:wmf100_bin_par}.

For these simulations core collapse occurs at around 20 N-body time
units (see Fig. \ref{fig:wmf100_struct_par} [upper left panel] and Table
\ref{tab:cc_par}), the wider mass function (compared to ``MF10'' runs)
further speeding up the evolution. Qualitatively both the
structural/energy parameters and the properties of the binaries compare
well with the standard case. Only the binary binding energies are again
higher as compared to the ``MF10'' runs, and the cumulative distribution
of the binding energies is steeper, biased to higher energies. In
general, the MF100 runs show larger scatter compared to the other runs.
This is due to the larger stochastic effects for the highest masses
caused by the wider mass range.

Both the binary properties and the structural/energy parameters are in
good agreement. Weakly significant inconsistency is found for the
half-mass radius and the potential energy evolution only. For these
runs, even the energy conservation is not inconsistent, as the main
differences occur only before core collapse, at ages which are largely
removed by skipping the first 10 N-body time units for the
bootstrapping.

For the MF100 test clusters we also test the eccentricity distribution
of unperturbed binaries (again for the test statistics clusters, hence
485 clusters instead of 62 clusters in the main simulations) against
various power-law relations and find a range in $\alpha$ = 1.9 -- 3.0
with Kuiper test probabilities $>$ 10\%, with a plateau of probability
$>$ 95\% for $\alpha$ = 2.3 -- 2.6. A thermal distribution has a Kuiper
test probability = 28.1\%, hence can not be rejected.

We again split the complete sample in thirds and employ Kuiper tests to
quantify the probability of the subsamples' eccentricity distribution
being thermal. Binaries are consistent with a thermal eccentricity
distribution for: small semi-major axes, high binding energies, high
primary mass, large distances from the cluster centre and (to a lesser
extend) large mass ratios. For each of those subsets when compared with
a thermal distribution a Kuiper test gives p-values $\ga$ 80 per cent.
Except for the small mass ratio subset (which is comparable with a
thermal distribution at p-value $\approx$ 40 per cent) for the opposite
subsets a thermal distribution is at least weakly significantly,
 if not more strongly, rejected.

\subsection{Using {\sc starlab} ``MF10 runs'' to test for possible
biases introduced by analysis procedure}
\label{sec:reswobin}

We checked whether the binary reconstruction with {\sc starlab/kira}
introduced spurious effects. 

The structural parameters are largely unaffected. Minor differences in
the binary parameters originate in the inclusion of perturbed binaries
into the sample before full binary reconstruction.

The main differences occur for the energies. Before binary
reconstruction, binaries are treated as 2 separate stars. Their orbital
velocities contribute therefore to the total kinetic energy of the
cluster. Their binding potential energy constitutes a significant part
of the clusters total potential energy. In the case of a fully
reconstructed binary tree structure both the orbital velocities' kinetic
energies and the binding potential energies are treated separately from
the total cluster values, leading to an apparent ``loss'' of total
energy.

However, the temporal parameter evolutions derived from {\sc 
starlab} vs {\sc nbody4} simulations show very similar bootstrap results
both before and after the binary reconstruction. We therefore conclude
that the binary reconstruction does not lead to systematical changes.

In addition, we checked whether the splitting into and analysis of
single snapshots introduces systematic differences. We pass the full
1000-snapshots {\sc starlab} output through the {\sc starlab} analysis
routine {\sc hsys\_stats} (like we do for the single snapshots) and
statistically compare the results with the results from the
single-snapshot approach. We find slight differences induced by the
resetting of the centre-of-mass for the single-snapshot approach, of
which none is statistically significant (except for the centre-of-mass
itself). Likewise, the binaries (both perturbed and unperturbed) do not
show significant differences. Alone the multiples' properties show
significant differences (due to the spurious reconstruction of a small
number of multiples far out of the cluster centre), and should be
treated with caution.

\section{Energy conservation}
\label{sec:resEC}

Ideally, the total energy should be constant during each simulation
(except for the locking up of energy in binaries), and hence also in the
median datasets. In practice, numerical inaccuracies lead to changes in
the total energy. This energy error can be seen e.g. as the change of
total energy per N-body time unit. These errors are larger if higher
accuracy and hence more timesteps per N-body time unit are required,
e.g. during core collapse and close encounters. 

After core collapse, the energy error rises steeply, indicating that
after core collapse, the errors are dominated by close encounters and
binaries (and more timesteps per N-body unit are required). Prior to core
collapse the errors originate from inaccuracies of the Hermite
integrator alone. For ages well after core collapse, the energy error
steadily decreases as the cluster expands and the number of timesteps
per N-body unit drops.

\begin{figure*}
\begin{center}
  \vspace{-0.5cm}
  \hspace{1.2cm}
  \begin{tabular}{cc}	
	\includegraphics[angle=270,width=0.4\linewidth]{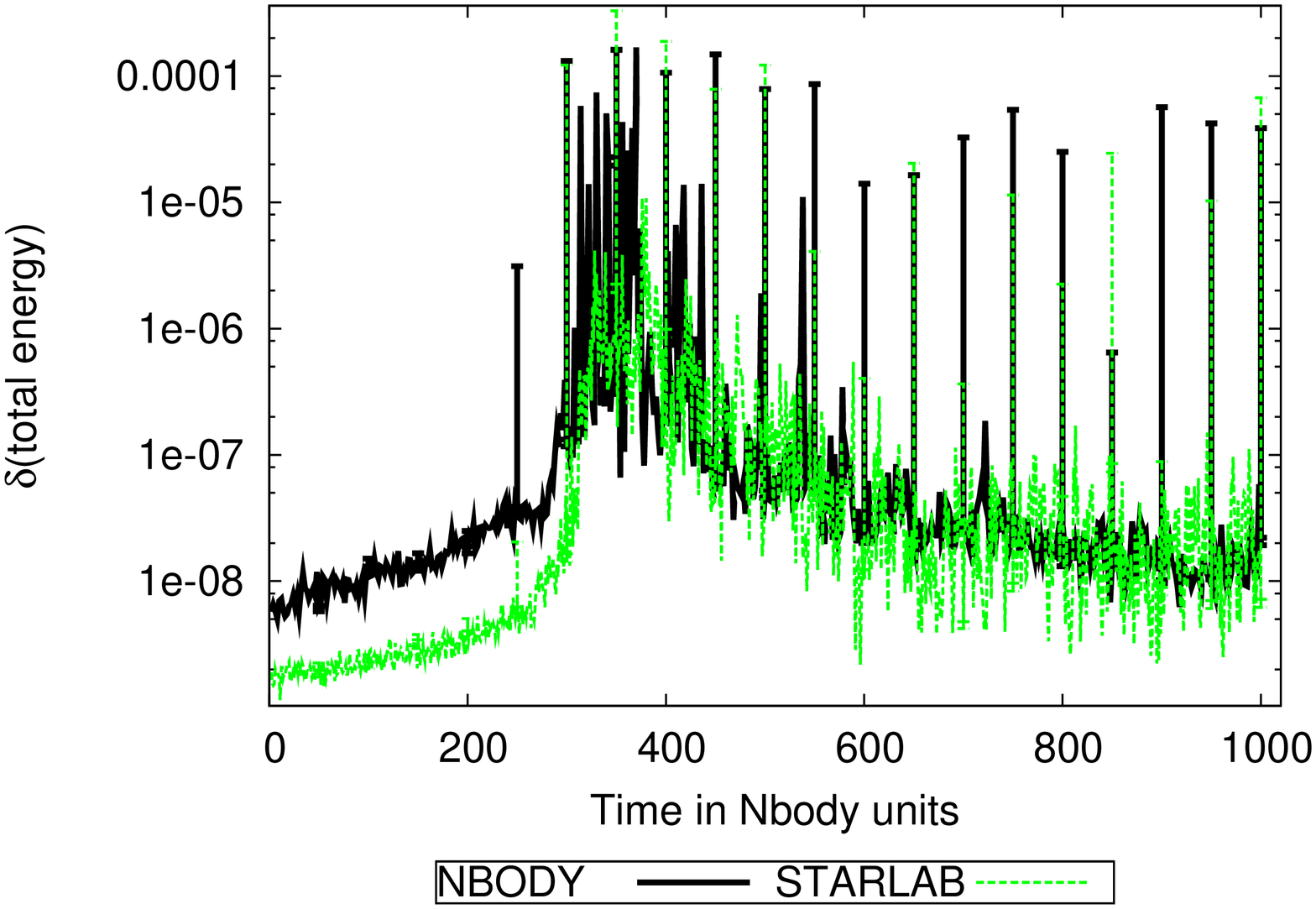} & 
	\includegraphics[angle=270,width=0.4\linewidth]{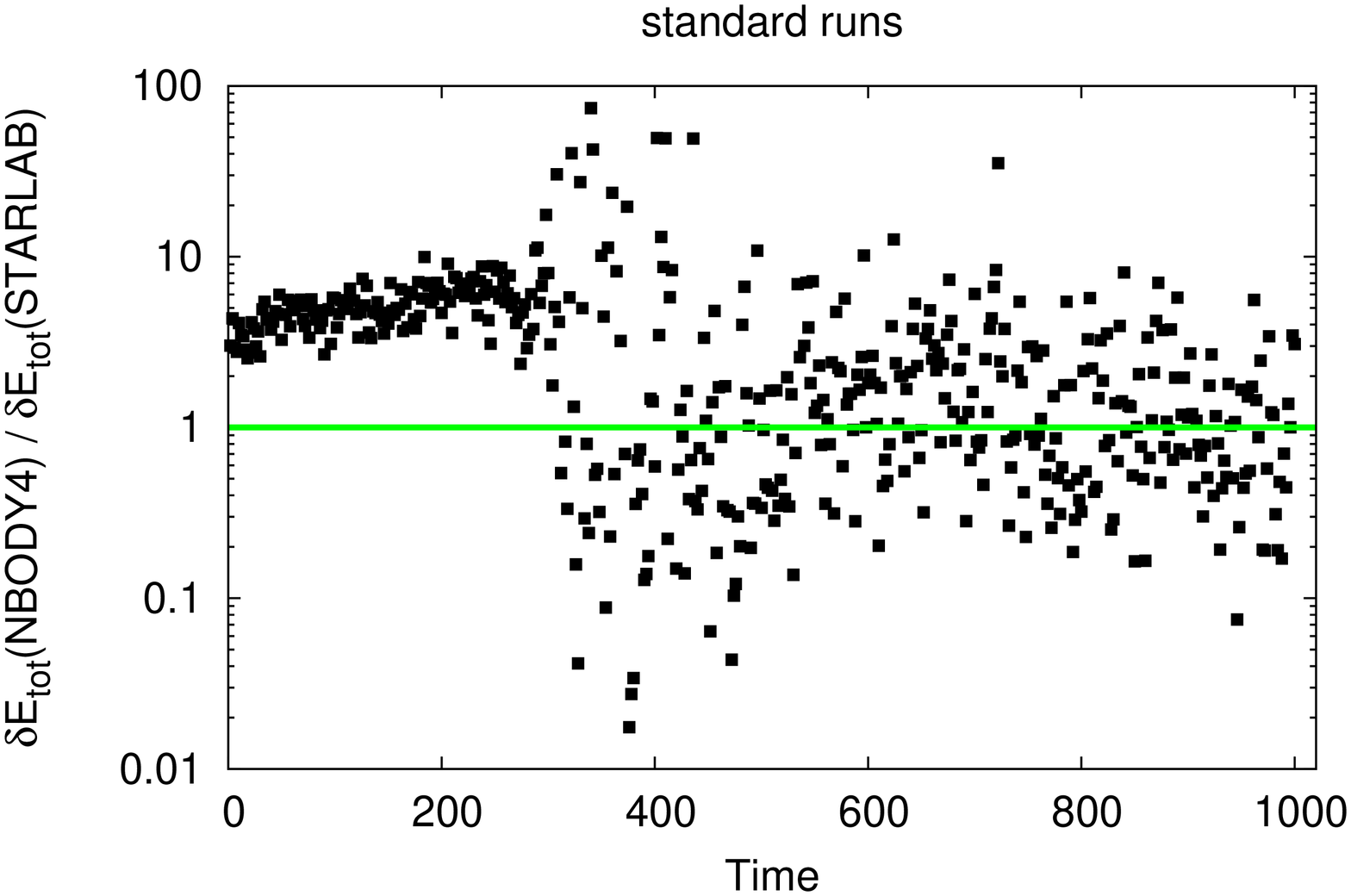} \\

	\includegraphics[angle=270,width=0.4\linewidth]{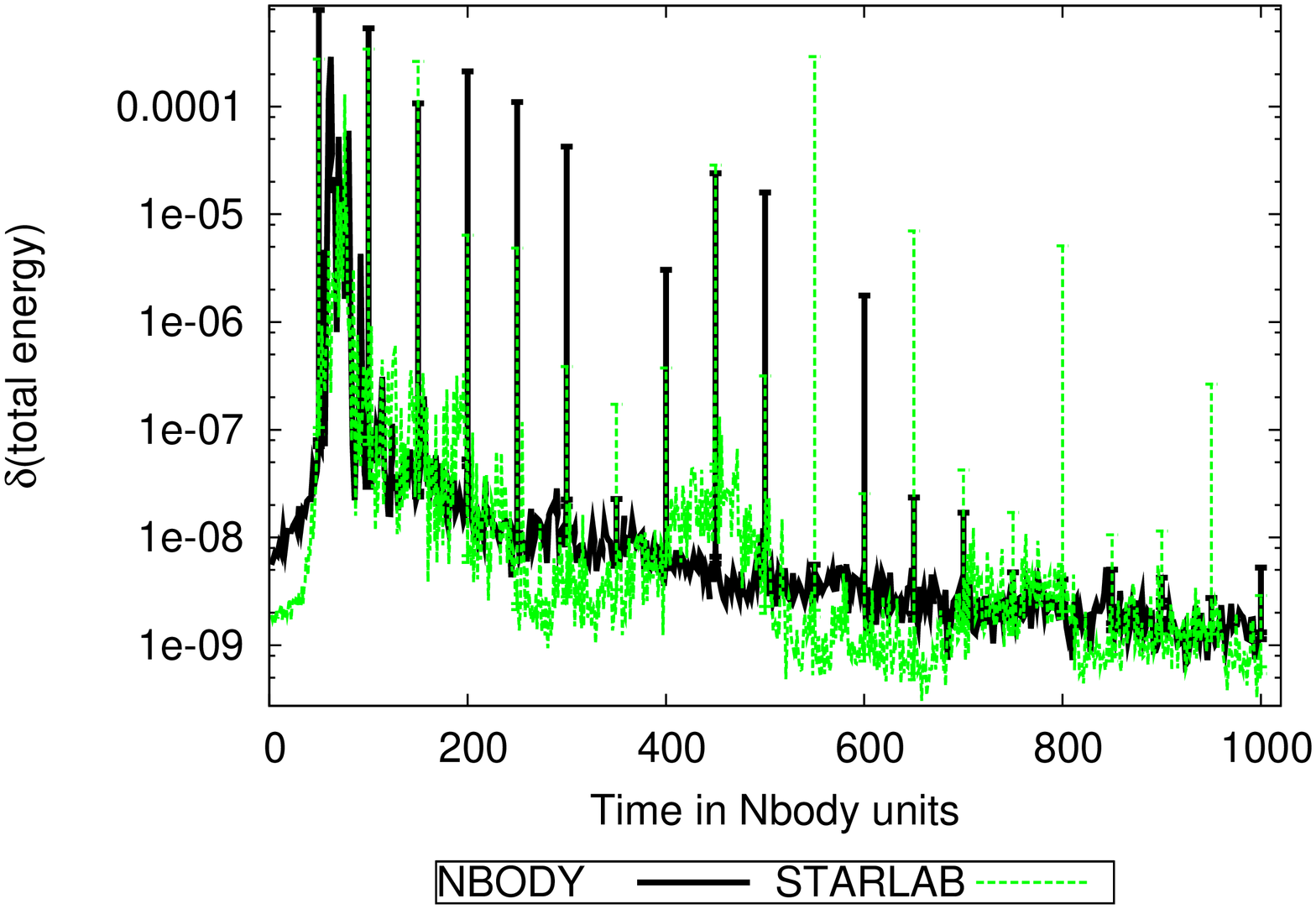} &
	\includegraphics[angle=270,width=0.4\linewidth]{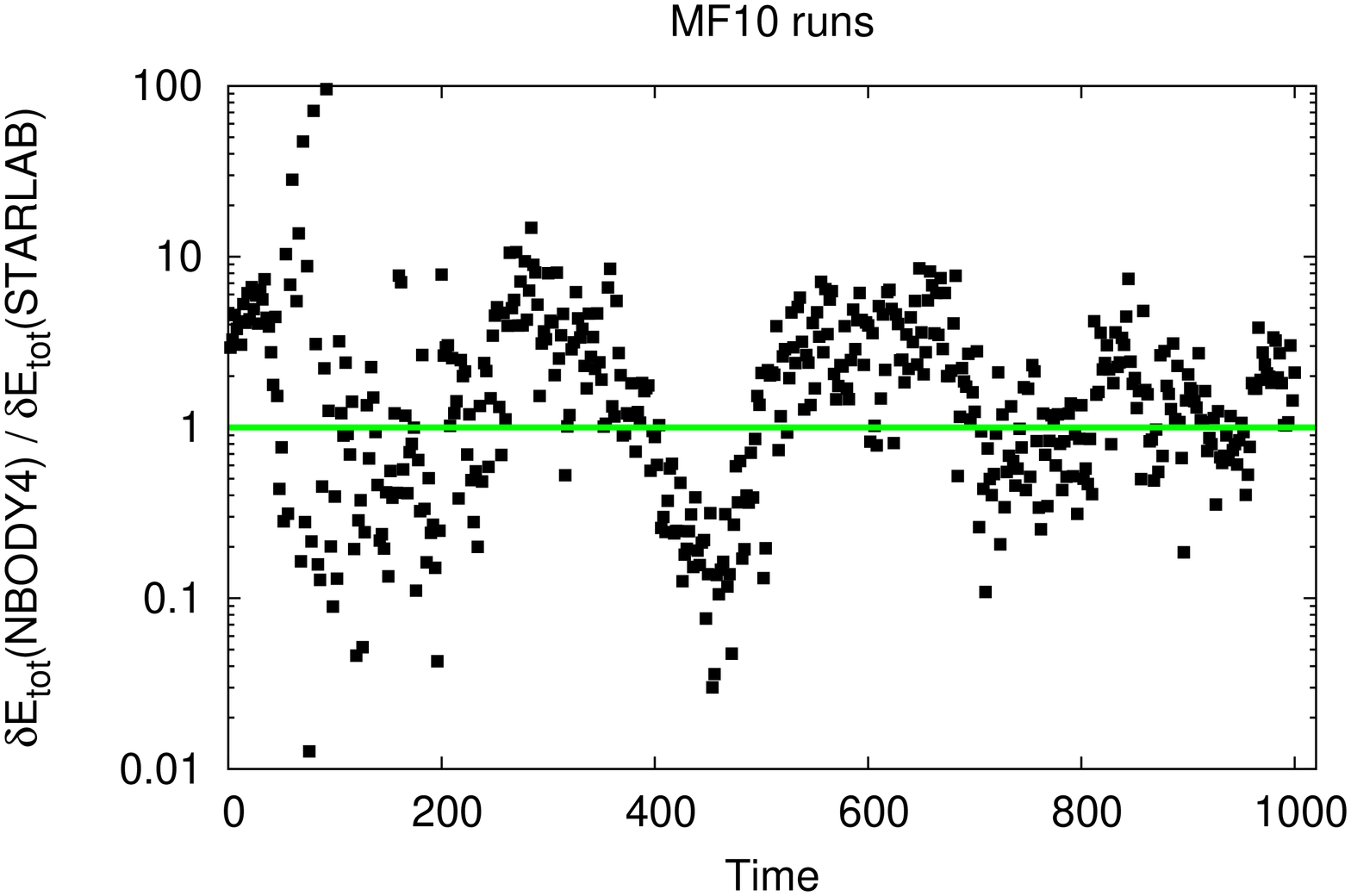} \\  

	\includegraphics[angle=270,width=0.4\linewidth]{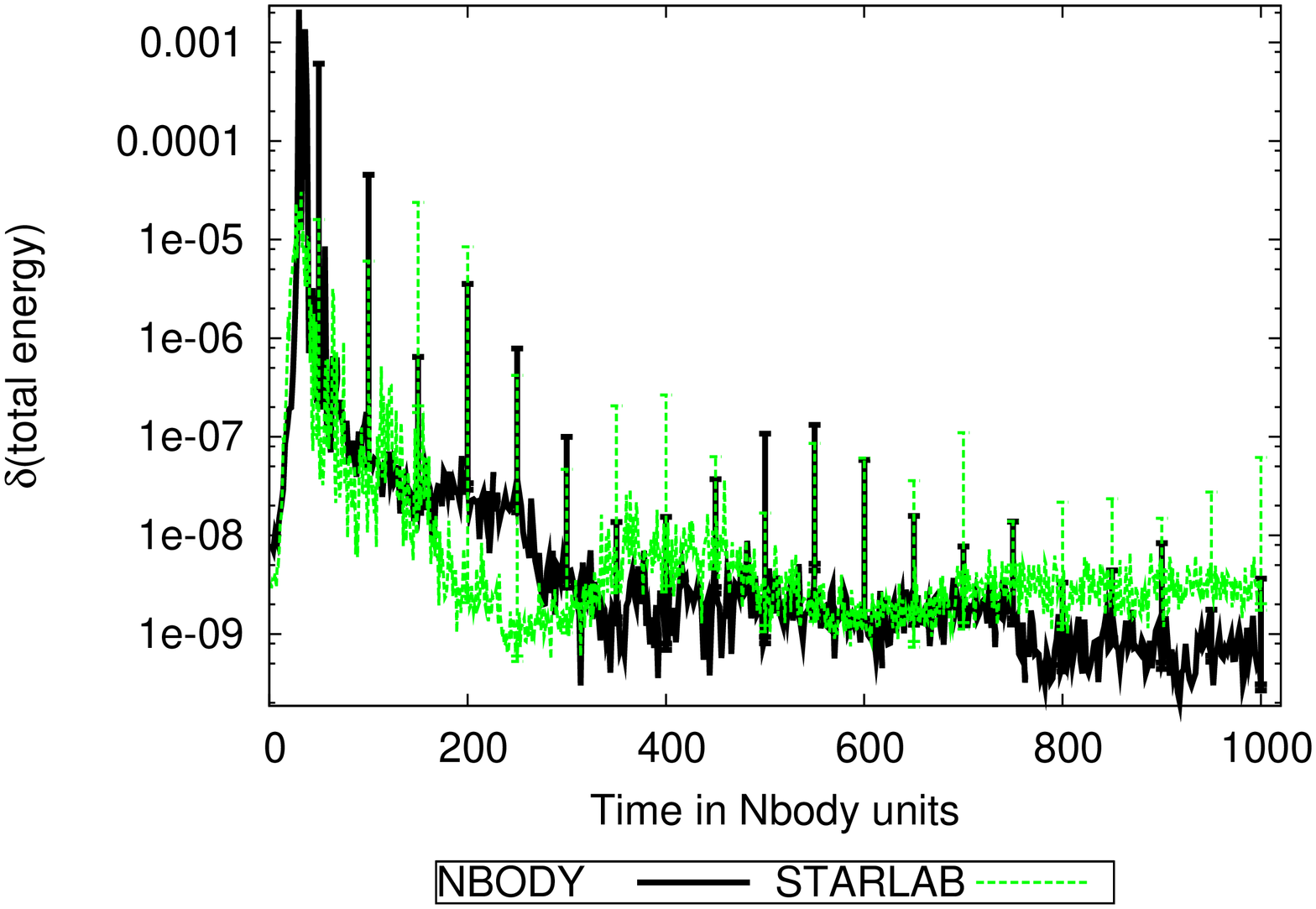} & 
	\includegraphics[angle=270,width=0.4\linewidth]{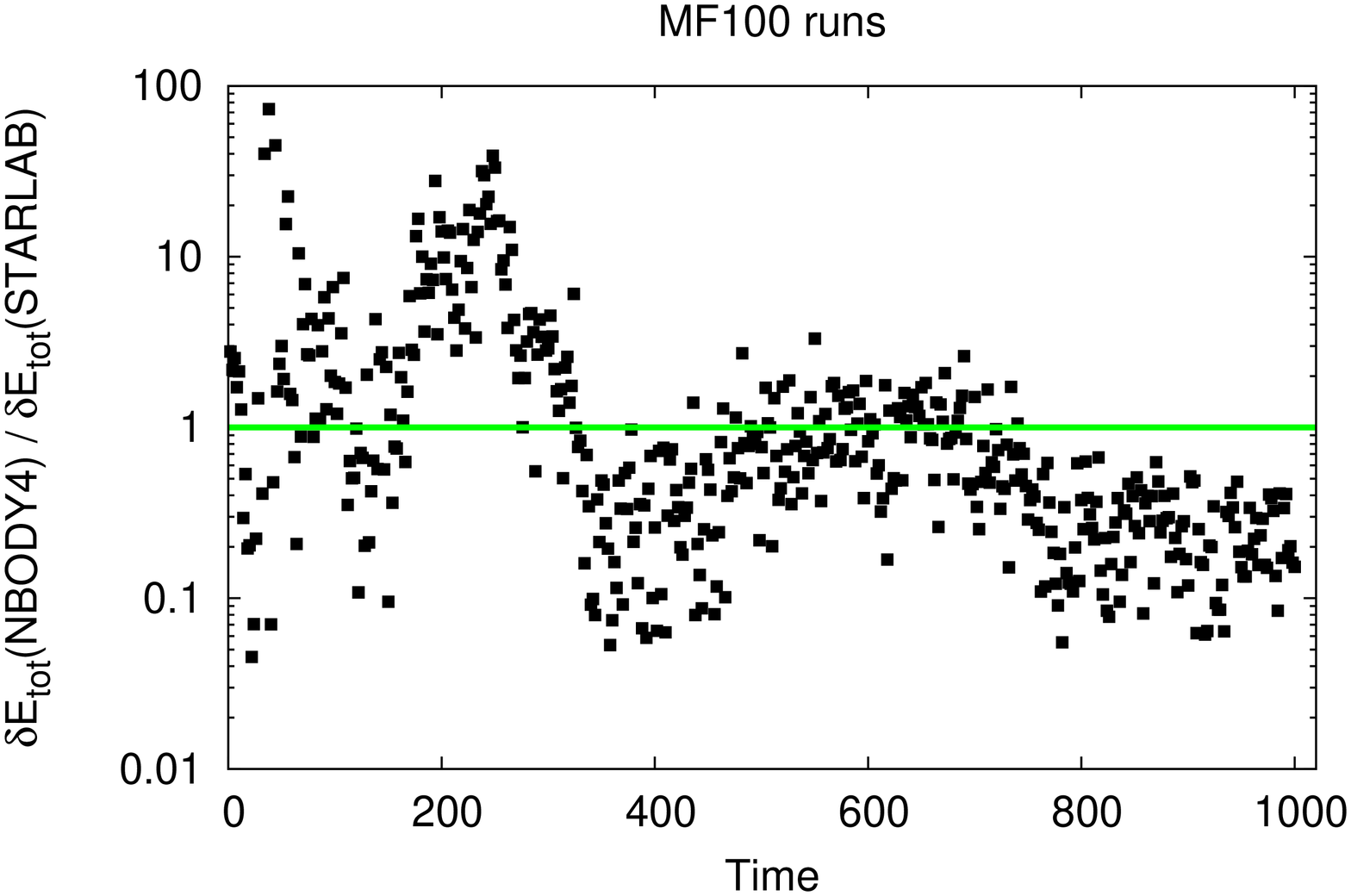} \\
  \end{tabular}
\end{center}

\caption{The energy conservation for  {\sc nbody4} (black lines) and
{\sc starlab} (green lines) for: ``standard runs'' (top), ``MF10
runs'' (middle), and ``MF100 runs'' (bottom). The  right
panels show the respective energy error ratios {\sc nbody4}/{\sc starlab}. A
horizontal lines as y=1 is added, to guide the eye.}
\label{fig:acc_acc}
\end{figure*}

The energy conservation from {\sc nbody4} shows roughly the same shape
as the one from {\sc starlab} (see Fig. \ref{fig:acc_acc}). However,
before core collapse the energy conservation from {\sc nbody4} is
systematically worse than from {\sc starlab}, roughly a factor 3 -- 10,
with median deviations of 4 -- 5 ($\sim$2 -- 3 for MF100). After core
collapse, the median deviations are for all sets of simulations well
below a factor 2. However, the overall scatter strongly increases (by
factors 30 -- 2000) due to the non-conservative numerical effects during
binary interactions. Interestingly, the {\sc starlab} data show similar
energy conservation compared to the {\sc nbody4} data, despite the much
more sophisticated and programming expensive regularisation treatment in
{\sc nbody4}.

Performing an integration with the KS routines switched off shows that
the larger energy errors of {\sc nbody4} prior to core collapse come
from the KS formalism (possibly from the interface between the main
integrator and the KS algorithm), although no details or solutions could
be found. However, the energy conservation in {\sc nbody4} is still
good, and clearly sufficient for most applications.

\begin{table*}

\caption{Results from the bootstrap test. Given are the fractions (in
\%) of the test distributions more deviating than the main simulations,
i.e. the smaller this number the less alike the distributions are.
For details see text.}

\begin{center}
\begin{tabular}{l c c c}
\hline
parameter & std & MF10 & MF100 \\
          & $\Delta_{12}$  $\Gamma_{12}$ & $\Delta_{12}$  $\Gamma_{12}$  & $\Delta_{12}$  $\Gamma_{12}$ \\
\hline
r$_{\rm core}$        &  10.79 \textcolor{blue}{\bf 7.84}  &  59.56 83.17  & 24.99 27.37 \\ 
r$_{\rm half}$        &  43.79 66.70                          &  36.39 56.13  & \textcolor{blue}{\bf 9.8} \textcolor{blue}{\bf 7.8}\\ 
r$_{\rm max}$         &  80.74 40.58                          &  21.77 25.85  & 25.26 26.73 \\
$<$mass$_{\rm core}>$ &  28.70 77.44                          &  79.56 76.81  & 92.18 83.38 \\
$\arrowvert$ density centre $\arrowvert$    &  94.25 99.41 &  95.24 58.26  &  91.26 94.32 \\ 
\hline
E$_{\rm pot}$         &  40.10 83.08 &  59.52 78.82           &    \textcolor{blue}{\bf 5.60} \textcolor{blue}{\bf 6.16} \\ 
E$_{\rm kin}$         &  32.43 39.08 &  48.54 56.38           &    68.97 77.52 \\ 
E$_{\rm tot}$         &  \textcolor{magenta}{\bf 0.86} \textcolor{magenta}{\bf 0.72}   &  45.64 47.75  &    80.35 98.17\\ 
$\delta$E$_{\rm tot}$ &  \textcolor{magenta}{\bf 0}    \textcolor{magenta}{\bf 0}      &  \textcolor{magenta}{\bf 0} \textcolor{magenta}{\bf 0}      &  58.13 99.18 \\ 
\hline
\end{tabular}
\label{tab:boot_results}
\end{center}
\end{table*}

\begin{table*}

\caption{Kuiper test results. Given is the probability (in \%) that for
the given setup the binary properties from {\sc starlab} vs. {\sc
nbody4} are drawn from the same distribution. Quantities are: axis =
semi-major axis, ecc = eccentricity, D1 = distance of the binary from
cluster centre, in units of the core radius, , D2 = distance of the
binary from cluster centre, in units of the half-mass radius, E/kT =
binding energy in E/kT, m2/m1 = mass ratio secondary mass / primary
mass. \#1 = number of binaries in the {\sc starlab} simulation, \#2 =
number of binaries in the {\sc nbody4} simulation. In addition,
e$^2_{SL}$  and e$^2_{NB}$ are the probabilities, that respectively the
{\sc starlab}/{\sc nbody4}  data for the eccentricity distributions are
compatible with a cumulative distribution of the form e$^2$ (equivalent
to a thermal eccentricity distribution $\sim$ 2*e). For a description of
the setups see text.} 

\begin{center}
\begin{tabular}{l c c  c c  c c c c c c c c}
\hline
setup & \#1 & \#2 &          & axis & ecc & D1 & D2 & E/kT & m2/m1 &          & e$^2_{SL}$ & e$^2_{NB}$\\
\hline
STD, unperturbed   & 333 & 322 &    & 39.7 & 99.5 & 42.7 & 76.4 & 66.2 & 100.0 &     &                         99.8 & 99.5 \\
STD, perturbed     & 31 & 24 &      & 91.8 & 99.5 & 96.3 & 99.9 & 88.0 & 100.0 &    &                          95.6 & 97.4 \\
STD, multiples     & 10 & 9 &       & \textcolor{blue}{\bf 9.6} & 99.4 & 85.0  & 99.8 & 20.7 & 100.0 &      &  99.9 & 99.9 \\
\hline
MF10, unperturbed  & 191 & 192 &    & 41.6 & 92.5 & 81.4 & 87.2 & 84.8 & 14.9  &      &                        29.3 & 82.2 \\
MF10, perturbed    & 50 & 37 &      & 94.8 & 60.4 & 58.8 & \textcolor{magenta}{\bf 3.3} & 80.6 & 50.3 &      & 99.5 & 99.1 \\  
MF10, multiples    & 14 & 15 &      & 33.9 & 57.4 & 32.2 & 37.4 & 37.4 & 30.5  &        &                      95.2 & 99.8 \\
\hline
MF100, unperturbed & 62 & 68 &      & 70.8 & 65.8 & 14.4 & 63.7 & 99.2 & 16.8  &        &                      95.1 & 94.7 \\
MF100, perturbed   & 36 & 44 &      & 17.3 & 98.5 & 77.7 & 95.8 & 24.7 & \textcolor{blue}{\bf 9.8} &        &  99.1 & 86.0 \\ 
MF100, multiples   & 29 & 33 &      & 81.1 & 52.2 & 39.9 & 83.7 & 68.5 & 91.6  &        &                      99.9 & 99.0 \\
\hline
\end{tabular}
\label{tab:kstest_bin}
\end{center}
\end{table*}


\section{Core collapse}
\label{sec:cc}

\begin{figure}
\begin{center}
	 \includegraphics[angle=270,width=0.8\linewidth]{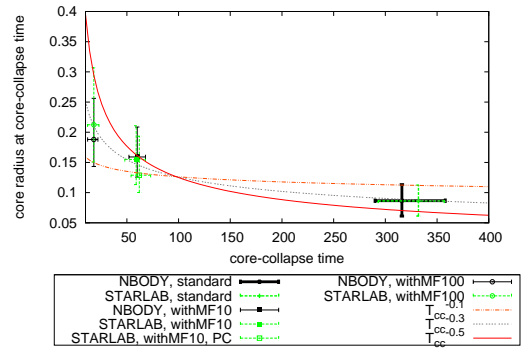}
\end{center}

\caption{Comparison of simulations using {\sc starlab} (green/grey) vs
{\sc nbody4} (black). Shown is the time at which core collapse occurs
(i.e. the time at which the first binary with more than 100 kT binding
energy is formed) vs the ``depth'' of core collapse (i.e. the  core
radius at core collapse time).}

\label{fig:all_cc_par}
\end{figure}

For the three settings investigated in this study (i.e. ``standard'',
``MF10'' and ``MF100''), the parameters of core collapse are presented
in Fig. \ref{fig:all_cc_par} and Table \ref{tab:cc_par}. For each
setting,  the values as derived from {\sc starlab} are consistent within
the 1$\sigma$ uncertainties with the values as derived from {\sc
nbody4}. 

We find a power-law dependence between time and ``depth'' of core
collapse, with $\alpha \sim$ -0.3 $\pm$ 0.2. This behaviour can be
understood qualitatively as follows: Core collapse is driven by the most
massive stars, which are sinking towards the centre of the cluster on a
timescale that is a fraction 1/M of the relaxation time. Hence core
collapse will occur faster in clusters with a broader mass spectrum. For
clusters with more massive stars, the core also reaches a state where
finite-N effects become important. At the same time core collapse is
halted at lower densities since energy generation due to binaries
becomes more efficient for higher mass binaries.

\begin{table}

\caption{Values of core collapse, i.e. at the time when the first
binary with binding energy $>$ 100 kT occurs. Given are the time of
core collapse t$_{cc}$, and the core radius at this time r$_{c,cc}$.
Shown are the median value from the individual runs, the uncertainty
ranges are the 16 and 84 quantiles, equivalent to 1$\sigma$ ranges of
the mean.} 

\begin{center}
\begin{tabular}{l l l}
\hline
setup & t$_{cc}$ & r$_{c,cc}$\\
\hline
\vspace{0.1in} STD, {\sc starlab} 	& 332$_{-39}^{+25}$ 	& 0.086$_{-0.025}^{+0.026}$\\
\vspace{0.1in} STD, {\sc nbody4} 	& 316$_{-26}^{+42}$ 	& 0.087$_{-0.026}^{+0.027}$\\
\hline
\vspace{0.1in} MF10, {\sc starlab} 	& 59$_{-11}^{+9}$ 	& 0.155$_{-0.042}^{+0.056}$\\
\vspace{0.1in} MF10, {\sc nbody4} 	& 60$_{-8}^{+8}$ 	& 0.159$_{-0.033}^{+0.050}$\\
\hline
\vspace{0.1in} MF100, {\sc starlab} & 18$_{-6}^{+5}$ 		& 0.212$_{-0.062}^{+0.095}$\\
               MF100, {\sc nbody4} 	& 18$_{-6}^{+4}$ 		& 0.188$_{-0.045}^{+0.068}$\\
\hline
\end{tabular}
\label{tab:cc_par}
\end{center}
\end{table}

\section{Escaping stars}
\label{sec:escapers}

\begin{figure*}
\begin{center}
  \begin{tabular}{cc}	
	 \includegraphics[angle=270,width=0.5\linewidth,viewport=80 20 500 700,clip]{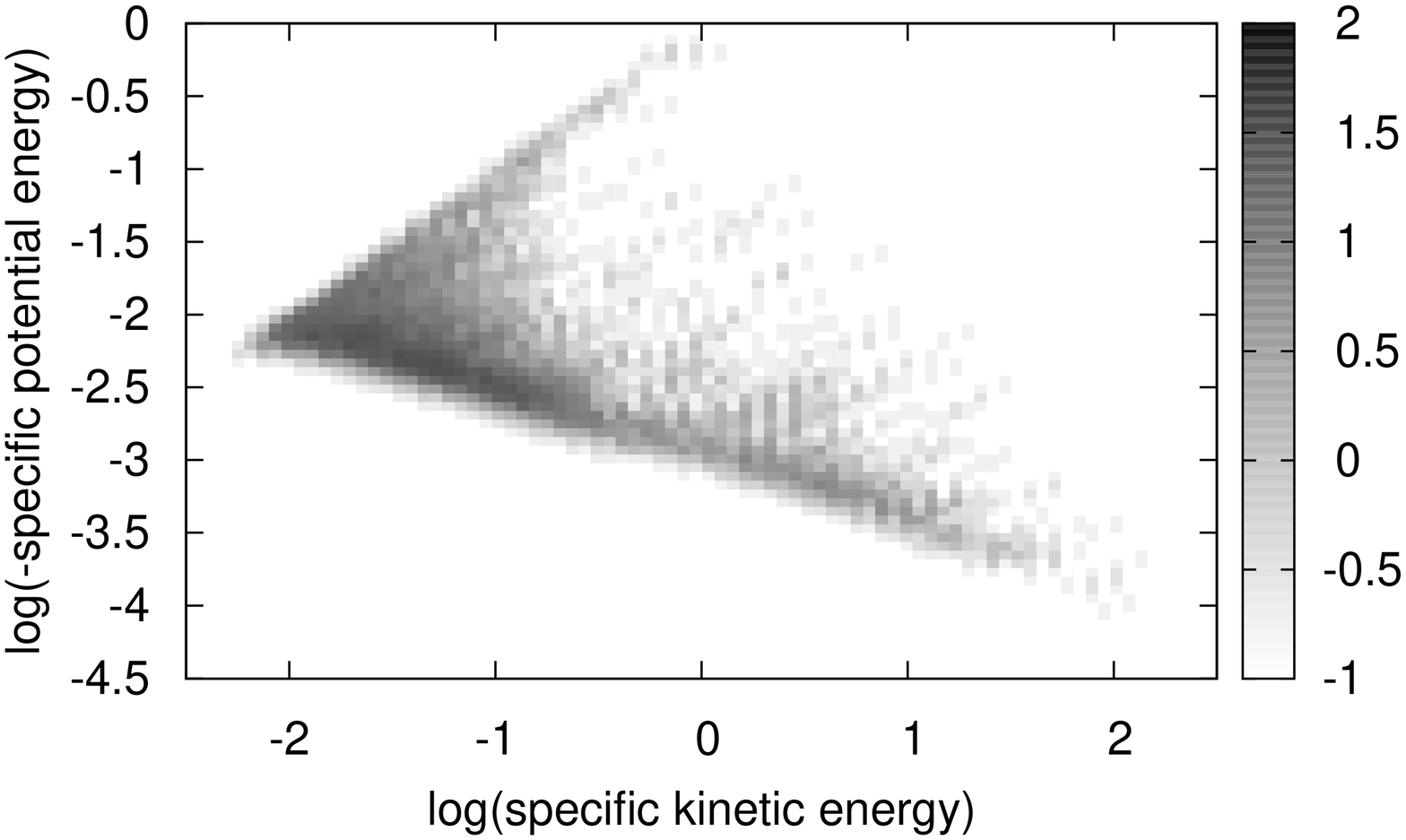} &
	 \includegraphics[angle=270,width=0.4\linewidth,viewport=20 00 500 700,clip]{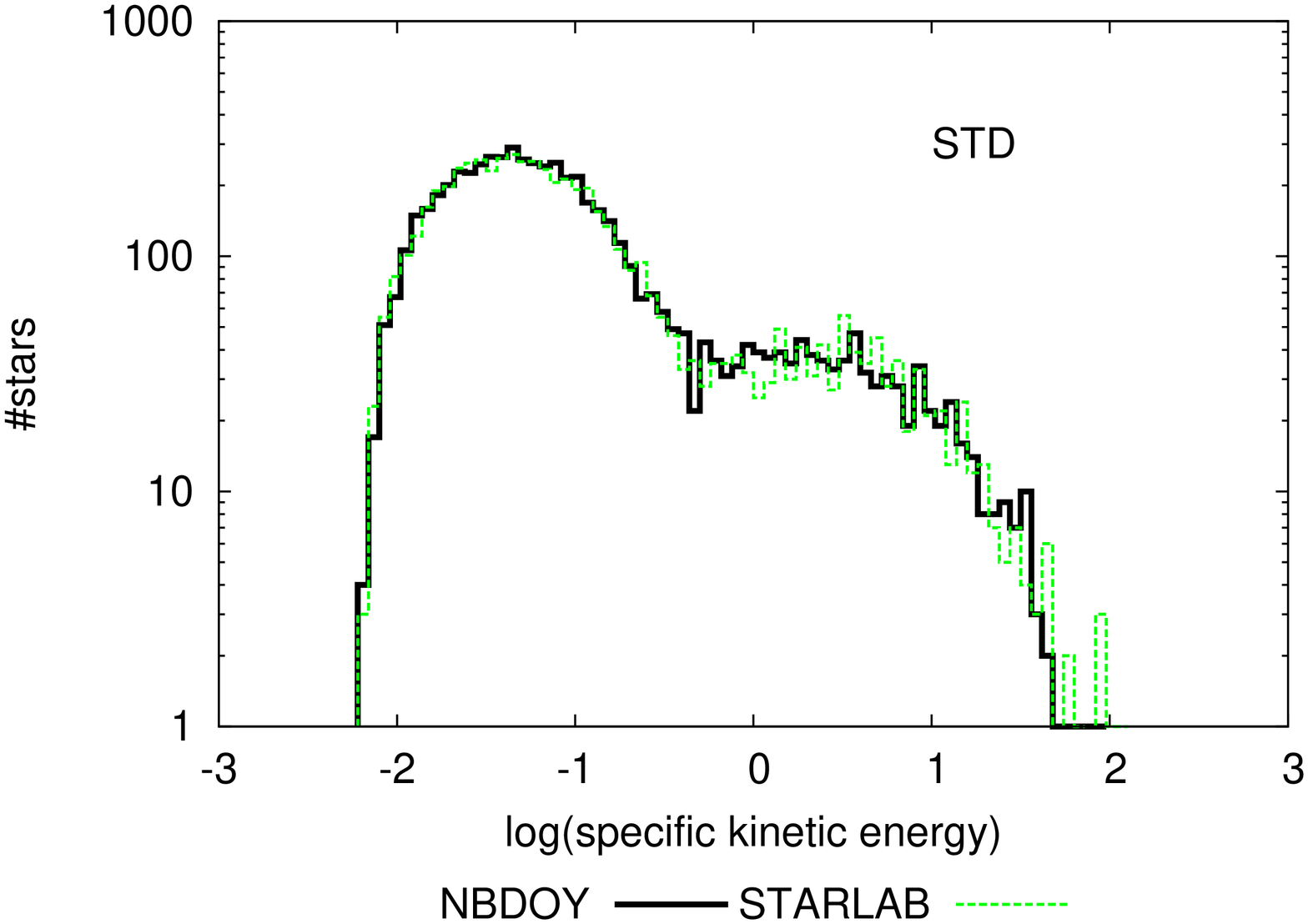} \\
	\includegraphics[angle=270,width=0.5\linewidth,viewport=80 20 500 700,clip]{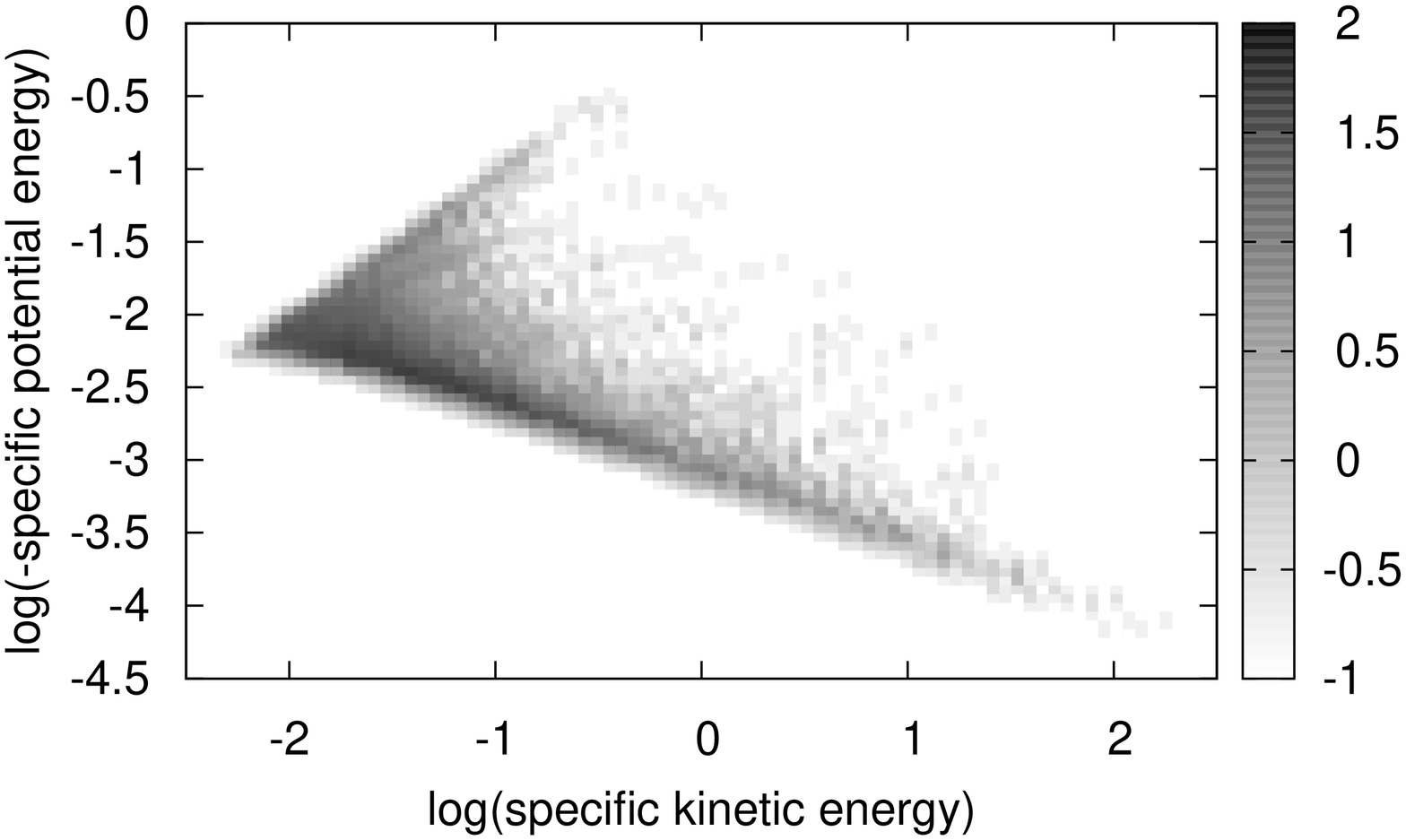} &
	\includegraphics[angle=270,width=0.4\linewidth,viewport=20 00 500 700,clip]{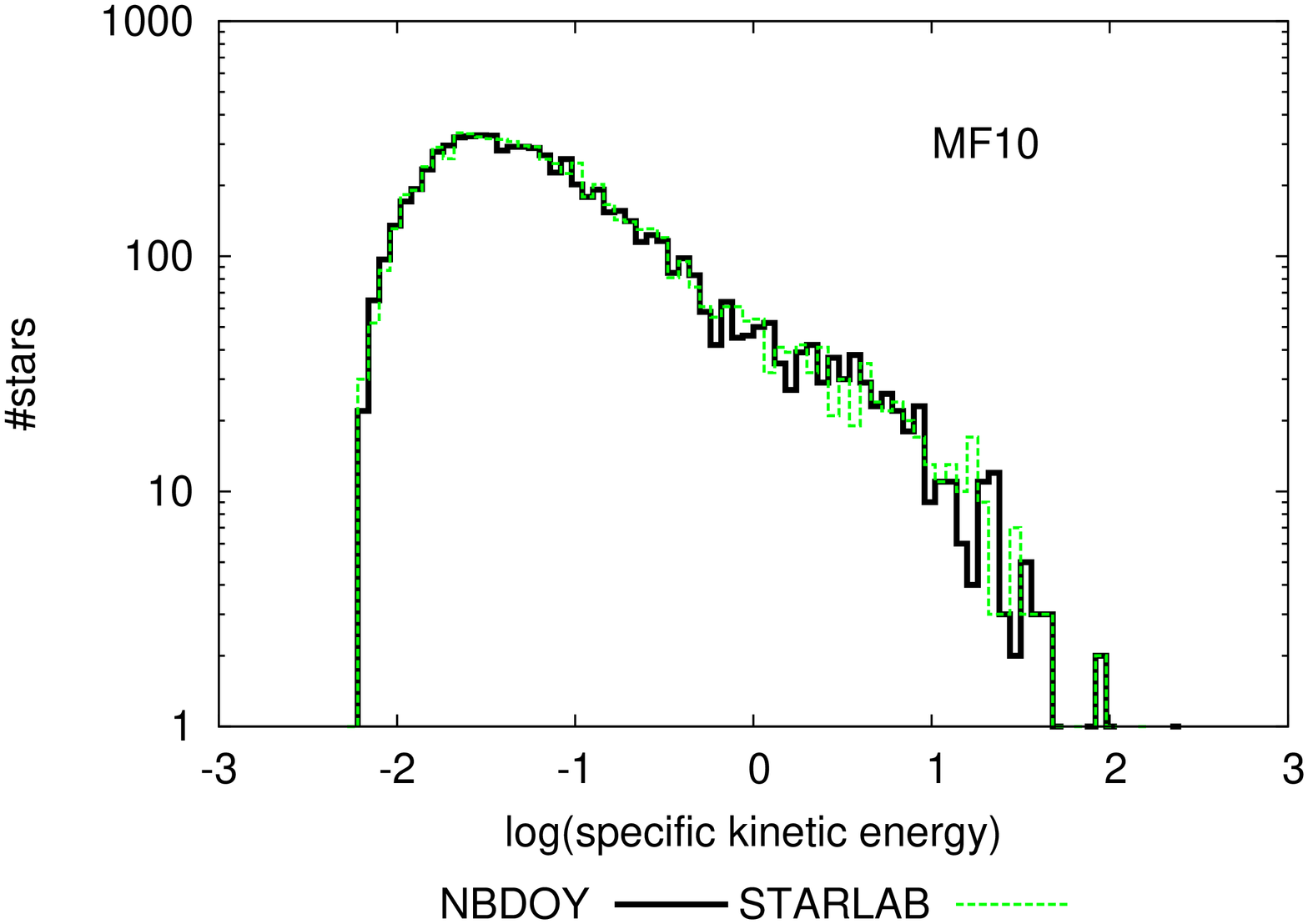} \\
	\includegraphics[angle=270,width=0.5\linewidth,viewport=80 20 500 700,clip]{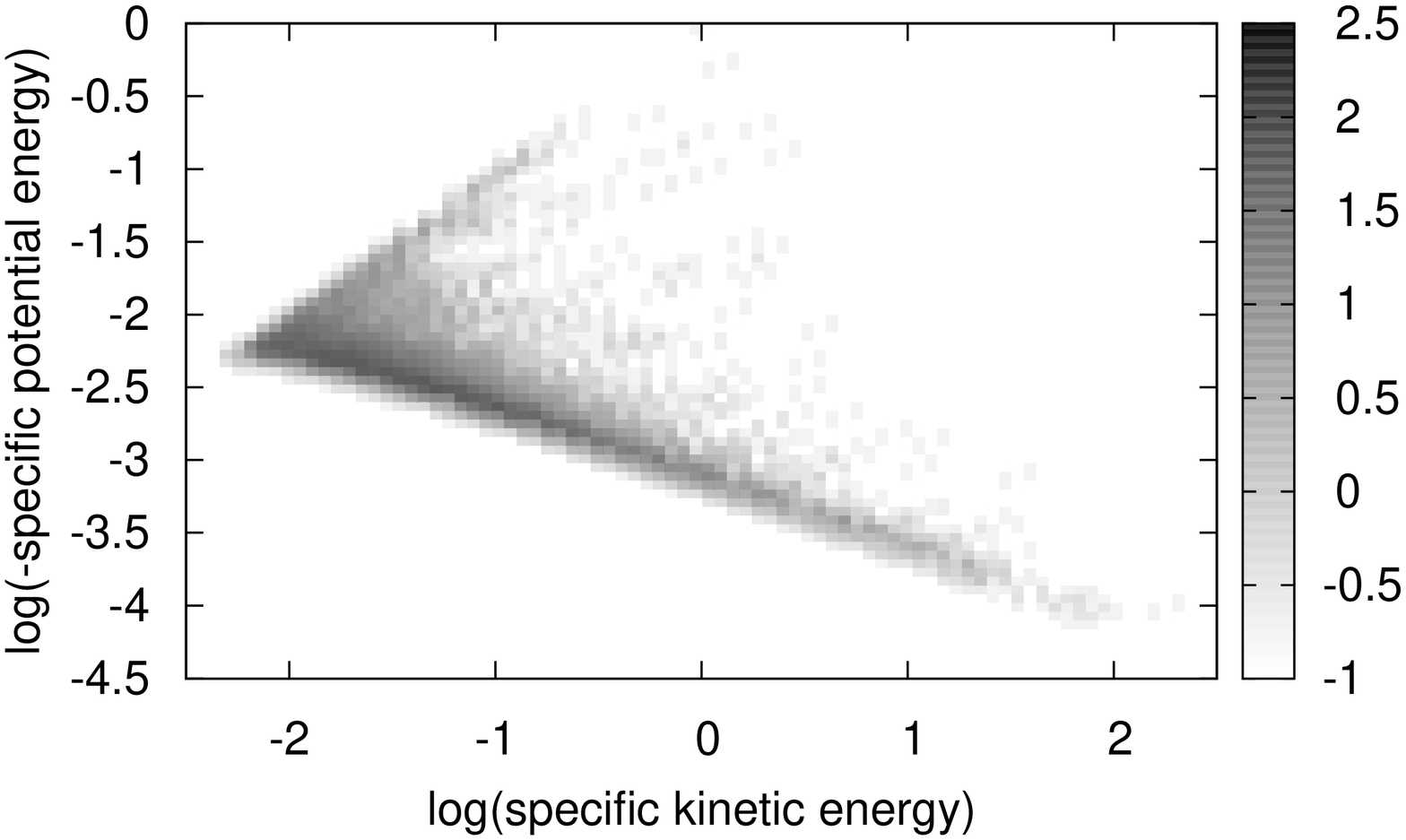} &
	\includegraphics[angle=270,width=0.4\linewidth,viewport=20 00 500 700,clip]{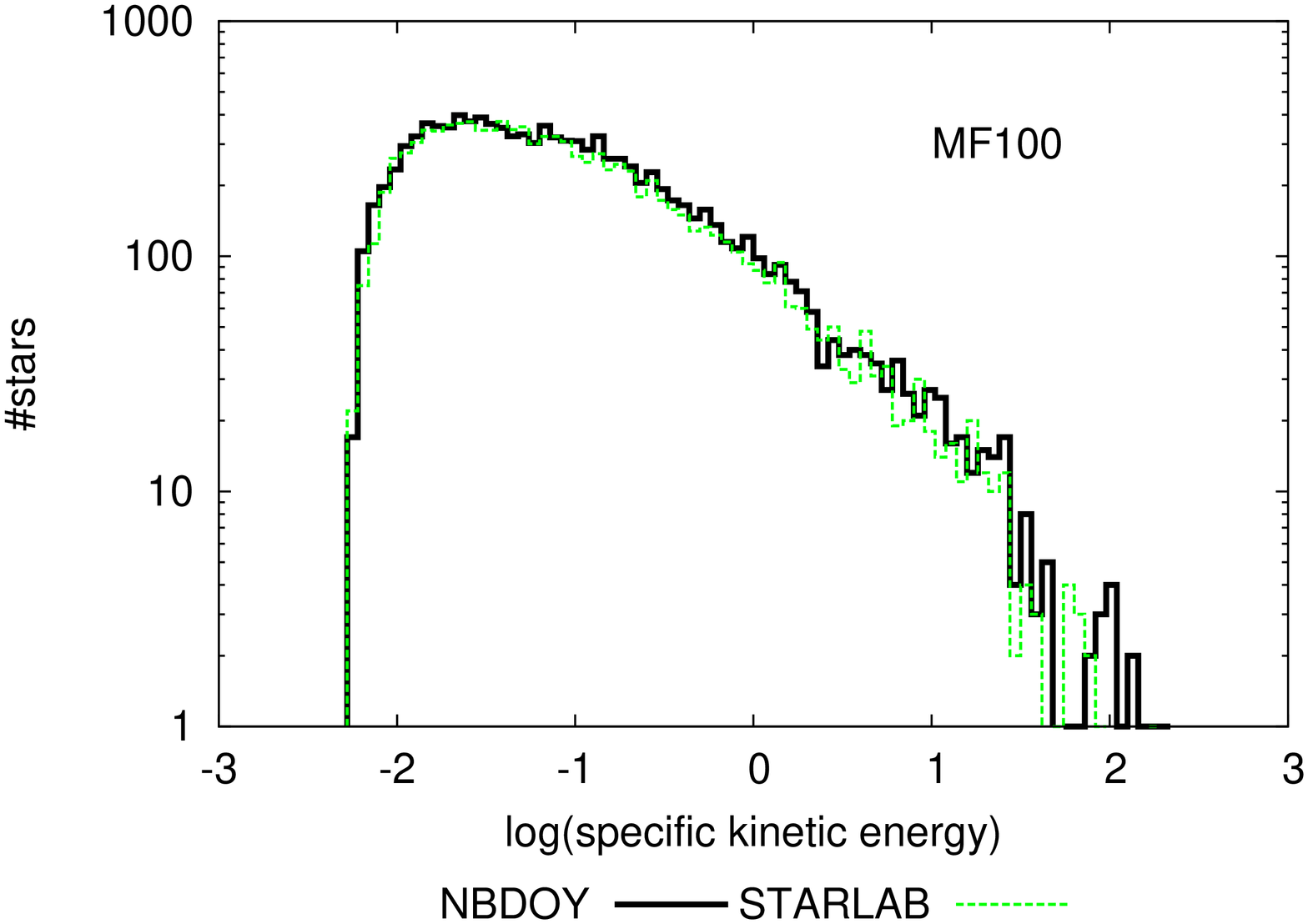} \\
  \end{tabular}
\end{center}

\caption{Specific kinetic and potential energy of escaping stars after
1000 N-body time units for standard setting (upper panels), MF10 runs
(middle panels) and MF100 (lower panels). Left panels: Shown is the
absolute number of stars in a given kinetic-vs.-potential energy bin
(logarithmic bins) for {\sc starlab} simulations (logarithmic grey
scaling). Right panels: Comparison of kinetic energy distributions from
{\sc nbody4} (black lines) and {\sc starlab} simulations (green lines).}

\label{fig:all_escaper}
\end{figure*}

As a final check we compare the specific potential and kinetic energies
of stars escaping the cluster, as especially their kinetic energies
might sensitively depend on details in the treatment of binaries and
close encounters in general. 

To this end we determine for each star in each simulation its potential
and kinetic energy from its position and velocity within the cluster,
and divide by the mass of the star. We choose to compare only data at an
age of 1000 N-body time units, hence sufficiently after core collapse.
Results are presented in Fig. \ref{fig:all_escaper}. 

The left panels depict the distribution of kinetic versus potential
energy from our {\sc starlab} runs for the different mass function
settings. A clear bifurcation is visible: the upper branch/edge consists
of stars which are barely unbound (and a fraction of those might become
recaptured by the cluster). With time  these barely unbound stars
migrate outwards in the cluster, into cluster  regions with low specific
potential energy (i.e. downwards in the left panels of Fig.
\ref{fig:all_escaper}). Stars with higher (specific) kinetic energy
migrate faster, hence in the same time reach farther distances (i.e.
regions characterised by lower specific potential energy). Core collapse
is the earliest and by far strongest event leading to the unbinding of
stars. Hence, the lower branches consist of stars which became unbound
during core collapse, as they had the longest time to travel farthest.
Stars in between the two main branches became unbound after core
collapse.

There are two main mechanisms to accelerate stars sufficiently to become
unbound: multiple weak encounters (``evaporating stars'') and single/few
strong encounters (``ejected stars'', usually star-binary or
binary-binary interactions). In the case of clusters with a stellar mass
function, ``ejected'' stars can originate either i) from a scattering
event of a single star on a binary (here the energy gain for the
scattered star is small, as the energy gain originates from the
shrinking binary orbit alone) or ii) from an exchange interaction (in
this case the energy gain for the ejected star is large, as the energy
gain originates from the orbit shrinking and the change in potential
energy, as primarily a low-mass binary component is exchanged by a
high-mass intruder star).

For an in-depth study of ``evaporating'' vs. ``ejected'' stars see
\citet{2008MNRAS.389..889K}. On average, ``ejected'' stars have higher
kinetic energies than ``evaporating'' stars. Especially the stars with
the highest kinetic energies are exclusively ``ejected'' stars. The
maximum velocity gain a star can get during an interaction with a binary
is of the order of the maximum orbital velocity of the binary stars. For
an equal-mass binary the orbital velocity is directly related to the
hardness of the binary, while for unequal-mass binaries the mass ratio
between the binary stars plays the dominant role. The hardening of a
binary is a long process. Hence, for equal-mass systems highly energetic
ejected stars can only appear well after core collapse (and the
formation of the first binaries), while for systems with a stellar mass
function they can appear already right at core collapse. This effect is
seen in Fig. \ref{fig:all_escaper}, upper left panel: for the standard
models, the lower branch shows a break at a specific kinetic energy
$\sim$ -0.3, where stars with higher kinetic energy are slightly offset
towards higher potential energies (i.e. closer to the cluster centre).
This can be understood if these high-energy ``ejected'' stars left the
cluster after the ``evaporating'' stars with lower kinetic energies (and
therefore had less travel time), due to the time required for a
sufficient hardening of the binaries. 

The distinction into ``evaporating'' and ``ejected'' stars can also
be seen in Fig. \ref{fig:all_escaper}, upper right panel, which show a
pronounced dip between these constituents. For systems with a stellar
mass function, the contribution from ``scattered'' stars increases for
wider mass functions, as the probability of a low-mass star being
scattered at a high-mass binary increases. This increases the
contribution of stars with intermediate kinetic energies, filling up the
dip seen in Fig. \ref{fig:all_escaper}, upper right panel. In addition,
the ``exchanged'' stars can get up to higher kinetic energies for wider
mass function, as the possible energy gain due to the change in
potential energy increases. This is seen at the high-energy end of the
distributions in Fig. \ref{fig:all_escaper}, right panels.

We employ again a Kuiper test to test for statistically significant
differences between the {\sc nbody4} and {\sc starlab} energy
distributions, both for the potential and the kinetic energy. For the
standard and the MF10 runs we find no statistically significant
deviations. For the MF100 runs, we find probabilities of 3.1\% (kinetic
energy) and 0.24\% (potential energy), hence significant/highly
significant deviations. Visual inspection shows that the energy
distributions are slightly narrower for the {\sc starlab} runs compared
to the {\sc nbody4} results. With $\sim$10,000 stars in each sample, the
Kuiper test gives a statistically significant difference. From a
more detailed analysis of the binned distributions shown in Fig.
\ref{fig:all_escaper}, we find deviations between the two distributions
of order $\la 2\sigma$ for $\sigma$ determined from the Poisson
distributions which are expected to describe the number of stars in each
bin.

\section{Conclusions}
\label{sec:conclusions}

We presented the first systematic in-depth study on how well results
from the two major N-body codes {\sc nbody} (here {\sc nbody4}) and {\sc
starlab} are comparable. We started with three sets of input models (50
initial configurations each, for three different stellar mass functions)
and evolved these input models independently with {\sc nbody4} and {\sc
starlab}. We analysed the results in a consistent way, and developed
statistical tools to quantitatively compare the median results of a
variety of parameters (for each stellar mass functions) derived using
the two codes.

Overall, the agreement between the results obtained from the {\sc
nbody4} runs and from the {\sc starlab} runs is very good. Statistically
significant deviations were only found for the energy conservation
before core collapse (where {\sc nbody4} is significantly worse, likely
due to problems at the interface between the main integrator and the KS
algorithm for close encounter treatment) and for the kinetic/potential
energy distributions of escaping stars in the MF100 runs (with the {\sc
starlab} distributions being slightly narrower).

While testing the binary eccentricity distributions against the common
assumption of a thermal distribution, we find good agreement for the
main simulations, both for {\sc starlab} and {\sc nbody4}. However,
extending the number of test clusters (for {\sc starlab} only), we find
statistically significant biases towards higher eccentricities than a
thermal distribution would predict. These deviations are driven by the
dynamically least evolved binaries, while stars with probably the
highest number of previous encounters tend to be more thermalised
(though especially for the MF10 setting [i.e. a narrow mass range]
statistically significant deviations remain for a number of subsets).

We tested our approach for biases, potentially induced by splitting the
simulations into single snapshots and by the binary tree reconstruction.
None of these effects result in statistically significant deviations.

In summary, we have shown that for purely dynamical N-body modelling
results obtained from {\sc starlab} and {\sc nbody4} are consistent with
each other, allowing to combine these results without introducing
systematic effects.

A similar study including stellar/binary evolution still needs to be
performed.

\section{Acknowledgements} 

PA acknowledges funding by NWO (grant 614.000.529) and the European
Union (Marie Curie EIF grant MEIF-CT-2006-041108). We would like to
thank the International Space Science Institute (ISSI) in Bern,
Switzerland, where parts of the data were analysed and parts of this
paper were written, for their hospitality and support. We would like to
acknowledge the lively and stimulating discussions at the MoDeST-8
meeting in Bad Honnef (organized among others by Pavel Kroupa),
especially with Sverre Aarseth and Peter Berczik, as well as with
Andreas K\"upper. PA would like to acknowledge fruitful technical
discussions with Ines Brott and Evghenii Gaburov. 

\bibliographystyle{aa}
\bibliography{NB_SL_Comp}

\end{document}